\documentclass[aps,prl,twocolumn,amsmath,amssymb,superscriptaddress,showpacs,longbibliography]{revtex4-1}

\usepackage[switch]{lineno}
\usepackage{amsmath}
\usepackage{amssymb}
\usepackage{empheq}
\usepackage[parfill]{parskip}
\usepackage[active]{srcltx}
\usepackage{color}
\usepackage{array}
\usepackage{booktabs}
\usepackage{amsfonts}
\usepackage{dsfont}
\usepackage{graphicx}
\usepackage{natbib}
\usepackage{xcolor}

\begin{document}

\setlength{\parindent}{0.5cm}

\title{Sync and swarm: solvable model of non-identical swarmalators}

\author{S. Yoon}
 \affiliation{Departamento de F\'\i sica da Universidade de Aveiro \& I3N, Campus Universit\'ario de Santiago, 3810-193 Aveiro, Portugal}

\author{K. P. O'Keeffe}
 \affiliation{Senseable City Lab, Massachusetts Institute of Technology, Cambridge, MA 02139}

\author{J. F. F. Mendes}
 \affiliation{Departamento de F\'\i sica da Universidade de Aveiro \& I3N, Campus Universit\'ario de Santiago, 3810-193 Aveiro, Portugal}

\author{A. V. Goltsev}
 \affiliation{Departamento de F\'\i sica da Universidade de Aveiro \& I3N, Campus Universit\'ario de Santiago, 3810-193 Aveiro, Portugal}

\begin{abstract}
We study a model of non-identical swarmalators, generalizations of phase oscillators that both sync in time and swarm in space. The model produces four collective states: asynchrony, sync clusters, vortex-like phase-waves, and a mixed state. These states occur in many real-world swarmalator systems such as biological microswimmers, chemical nanomotors, and groups of drones. A generalized Ott-Antonsen ansatz provides the first analytic description of these states and conditions for their existence. We show how this approach may be used in studies of active matter and related disciplines.
\end{abstract}

\maketitle


Synchronization is a universal phenomenon \cite{winfree2001geometry, kuramoto2003chemical,pikovsky2003synchronization} seen in coupled lasers \cite{jiang1993numerical} and beating heart cells \cite{peskin75}. When in sync, the units of such systems align the rhythms of their oscillations, but do not move through space. Swarming, as in flocks of birds \cite{bialek2012statistical} or schools of fish \cite{katz2011inferring}, is a sister effect where the roles of space and time are swapped. The units coordinate their movements in space, but do not synchronize an internal oscillation.

The units of some systems coordinate themselves in both space and time concurrently. Japanese tree frogs sync their courting calls as they form packs to attract mates \cite{aihara2014spatio,ota2020interaction}. Starfish embryos sync their genetic cycles with their movements  creating exotic `living crystals' \cite{tan2021development}. Janus particles  \cite{yan2012linking,yan2015rotating,hwang2020cooperative}, Quincke rollers \cite{zhang2020reconfigurable,bricard2015emergent,zhang2021persistence}, and other driven colloids \cite{manna2021chemical,li2018spatiotemporal,chaudhary2014reconfigurable,zhou2020coordinating} lock their rotations as they self-assemble in space. The emergent `sync-selected' structures have great applied power. They have been used to degrade pollutants \cite{ursobreaking,dai2021solution,vikrant2021metal,tesavr2022autonomous}, repair electrical circuits \cite{li2015self}, and to shatter blood clots \cite{cheng2014acceleration,manamanchaiyaporn2021molecular}.

Theoretical studies of systems which mix sync with swarming are on the rise \cite{ventejou2021susceptibility,o2017oscillators,ha2019emergent,tanaka2007general,liu2021activity}. Tanaka {\it et al.} derived a universal model of chemotactic oscillators with diverse behavior \cite{tanaka2007general, iwasa2010dimensionality}. Active matter researchers studied a Vicsek model with self-rotating (synchronizable) units \cite{levis2019activity,ventejou2021susceptibility,liebchen2017collective} which imitate various types of colloid. O'Keeffe {\it et al.} introduced a model of 'swarmalators' \cite{o2017oscillators}, whose states have been found in the lab and in nature \cite{barcis2019robots,barcis2020sandsbots,zhang2020reconfigurable}, and is being further studied \cite{lee2021collective,hong2018active,lizarraga2020synchronization,o2018ring,ha2021mean,o2022swarmalators,sar2022swarmalators,o2019review,schilcher2021swarmalators}. 

Analytic results on swarmalators are sparse. Order parameters, bifurcations, {\it etc.} are hard to compute given the systems' nonlinearities and numerous degrees of freedom. Active matter such as the driven colloids mentioned earlier (which may be considered swarmalators) hard to analyze for the same reasons. The Vicsek model \cite{vicsek1995novel}, for example, requires an in-depth use of statistical physics tools (dynamical renormalization groups \textit{etc}) to be solved \cite{toner1998flocks}. As for generalized Vicsek models, often only the stability of the simple incoherent state is analyzed, while order parameters are found purely numerically \cite{levis2019activity,liebchen2017collective,ventejou2021susceptibility,liu2021activity}. As such, easily and exactly solvable models of active matter are somewhat rare.

This Letter shows how this gap in active matter and swarmalator research may begin to be closed using technology from sync studies. We use Kuramoto's classic self-consistency analysis \cite{kuramoto2003chemical} in hand with a generalized Ott-Antonsen ansatz \cite{ott2008low} -- two breakthrough tools -- to study swarmalators which run on a 1D ring. This simple model captures the essential aspects of real-world swarmalators/active matter, yet is also solvable: Its order parameters and collective states may be characterized exactly. To our knowledge, exact results for the order parameters of an active matter collective are few; in this sense our work contributes to this vibrant field.

\textit{Model.}--- The model we study is \cite{o2022collective}
\begin{eqnarray}
\dot{x}_i=v_i+\frac{J}{N} \sum_{j=1}^N \sin(x_j - x_i) \cos(\theta_j-\theta_i),
\label{eq: 1} \\
\dot{\theta}_i =  \omega_i+\frac{K}{N} \sum_{j=1}^N \sin(\theta_j-\theta_i) \cos(x_j - x_i),
\label{eq: 2}
\end{eqnarray}
where $(x_i, \theta_i) \in (S^1, S^1)$ are the position and phase of the $i$-th swarmalator and ($\nu_i, \omega_i$), $(J,K)$ are the associated natural frequencies and couplings. The $v_i, \omega_i$ are drawn 
from a Lorentzian distribution, $g_{v(\omega)}(x)=\Delta_{v(\omega)}/[\pi (x^2 + \Delta_{{v(\omega)}}^2)]$, with spreads $\Delta_v$, $\Delta_\omega$ and mean set to zero via a change of frame.

\begin{figure}[htpb!]
\centering
\includegraphics[width=8.5cm,angle=0.]{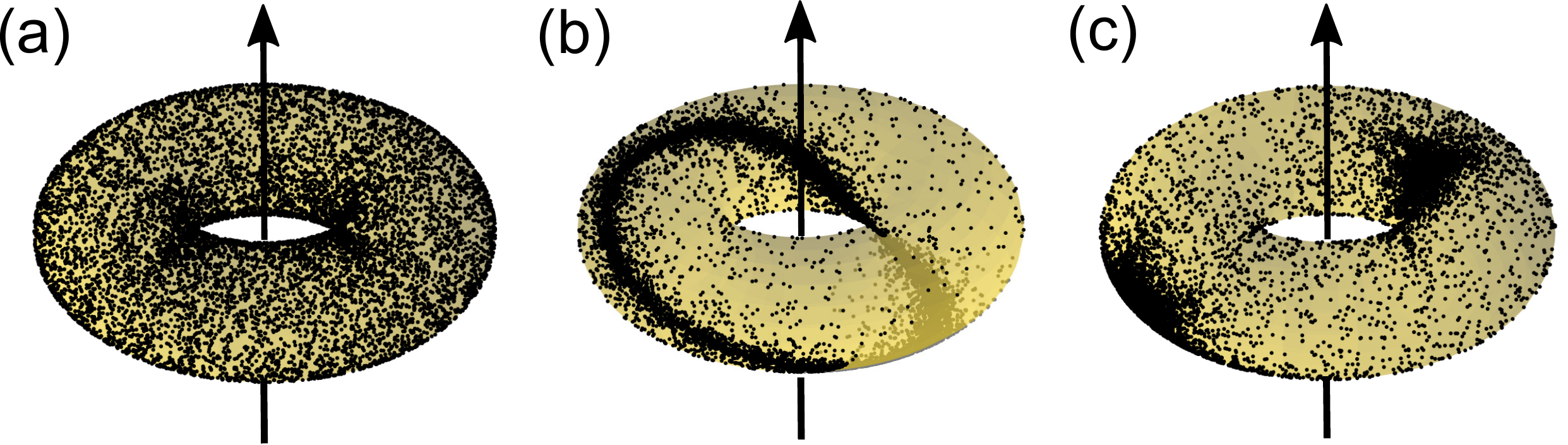}
\caption{Steady states of swarmalators (black dots) projected onto the unit torus. Data were generated by integrating Eqs.\eqref{eq: 1},\eqref{eq: 2} with an RK45 solver for $T = 500$ time units with adaptive stepsize for $N = 10^4$ swarmalators with $\Delta_{\nu} = \Delta_{\omega} = 1$. (a) Async state for $(J,K) = (1,1)$ where swarmalators are uniformly distributed in both space and phase. (b) Phase wave state for $(J,K) = (1,40)$ where positions and phases of swarmalators are correlated. (c) Sync state for $(J,K) = (8,9)$, where clusters of swarmalators synced in both space and time coexists with drifting swarmalators.}
\label{fig:torus}
\end{figure}

The phase dynamics Eq.~\eqref{eq: 2} are a generalized Kuramoto model where   now depends on their pairwise distance $K_{ij} = K \cos(x_j - x_i)$ \footnote{the original Kuramoto model has `all-to-all' coupling $K_{ij} = K$}. So for $K > 0$ neighbouring swarmalators synchronize more quickly than remote ones (the opposite occurs for $K < 0$). To treat sync and swarming on the same footing, the space dynamics Eq.~\eqref{eq: 1} are identical to Eq.~\eqref{eq: 2} but with $x_i$ and $\theta_i$ switched. Thus for $J>0$ synchronized swarmalator's swarm (in the sense of aggregating) more readily than desynchronized ones (the opposite for $J < 0$). In short, the equations model location-dependent synchronization, and phase-dependent aggregation. One can also think of them as sync on the unit torus (Fig.~\ref{fig:torus}) or as the rotational piece of the 2D swarmalator model \cite{SM}.

Introducing the variables
\begin{equation}
\zeta_i \equiv x_i + \theta_i, \,\,\,\,\,\,\,\,\,\, \eta_i \equiv x_i - \theta_i,
\label{eq: 5}
\end{equation}
let us write Eqs. (\ref{eq: 1}) and (\ref{eq: 2}) as a pair of linearly coupled Kuramoto models \cite{o2022collective}
\begin{eqnarray}
\!\!\!\!\!\dot{\zeta}_i &=&  v_i {+} \omega_i {-} J_{+}S_{+} \sin(\zeta_i {-} \Phi_{+}) {-} J_{-}S_{-} \sin(\eta_i {-} \Phi_{-}),
\label{eq: 8} \\
\!\!\!\!\!\dot{\eta}_i &=& v_i {-} \omega_i {-} J_{-} S_{+} \sin(\zeta_i {-} \Phi_{+}) {-} J_{+}S_{-} \sin(\eta_i {-} \Phi_{-}),
\label{eq: 9}
\end{eqnarray}
where $J_{\pm}\equiv(J \pm K)/2$ and 
\begin{equation}
W_{\pm} {\equiv}  \frac{1}{N}\sum_{j=1}^N e^{i(x_j \pm \theta_j)} = S_{\pm} e^{i\Phi_{\pm}}.
\label{eq: 7}
\end{equation}
These new order parameters measure the systems' space-phase order. When there is perfect correlation between space and phase $x_i = \pm \theta_i + C$, $S_{\pm} = 1$. When $x_i$ and $\theta_i$ are uncorrelated, $S_{\pm} = 0$. In a general case ($J \neq K$), the swarmalator and Kuramoto models belong to different classes of collective behavior. The coupling dependence on $S_{\pm}$ in Eqs. (4) and (5) leads to new collective states such as a mixed state in which $S_+$  and $S_-$ coexist. This state has no analogy in the Kuramoto model.

\begin{figure}[htpb!]
\centering
\includegraphics[width=8.5cm,angle=0.]{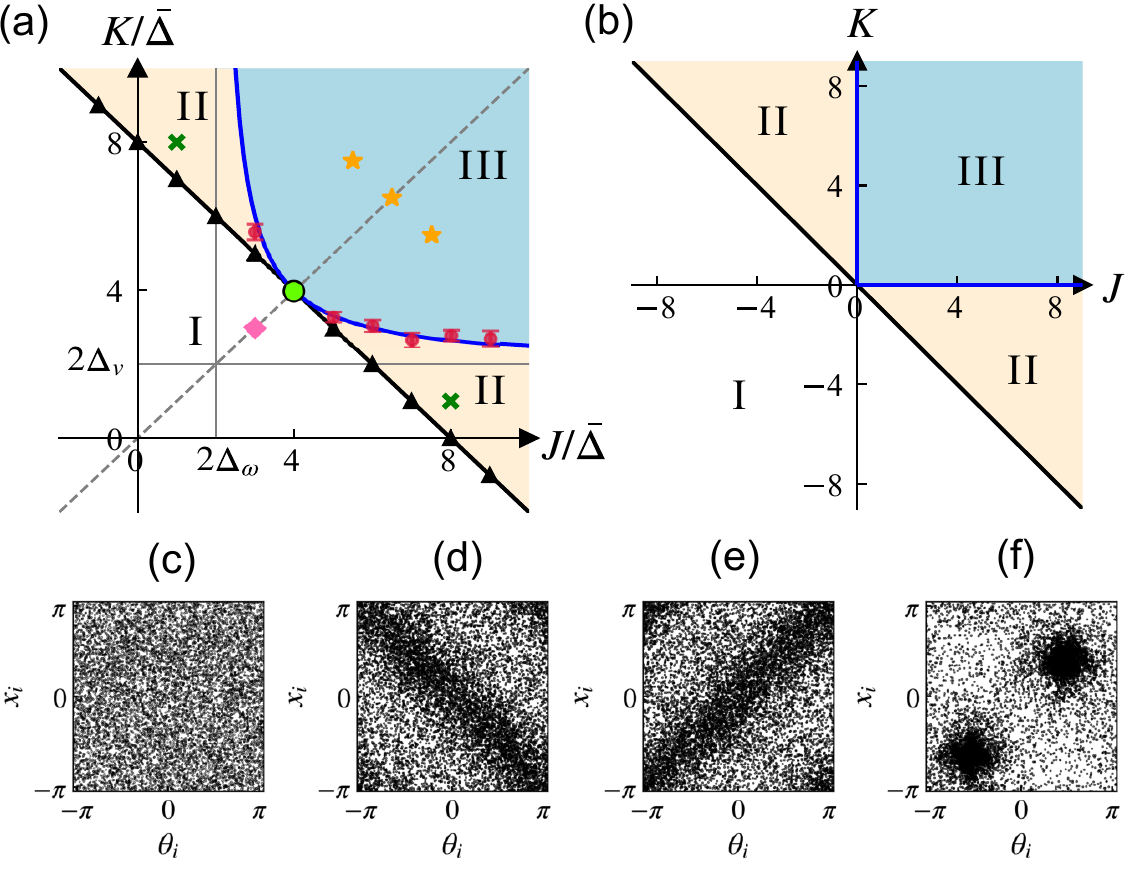}
\caption{
(a) Phase diagram of the swarmalator model in the $(J,K)$ plane (in units of $\overline{\Delta}=(\Delta_v+\Delta_{\omega})/2$). 
Regions I, II, III, and IV correspond to the $(0,0)$ (async), $(S,0)$/$(0,S)$ (phase wave), $(S_{1},S_{2})$ (mixed), and $(S,S)$ (sync) states. The black and blue solid lines represent the critical lines Eqs. (\ref{eq: 29}) and (\ref{eq: 43}). The purple solid line represents the critical line Eq.~(S69) in \cite{SM}. The black dashed line describes $J=K$ and the green circle is the 
tetracritical point. Symbols, 
black triangles, purple dots, and red circles are critical points found in simulations for $N=10^4$, $T=200$ with adaptive time step RK45 solver, and averaged by 20 realizations. (b) Phase diagram of the model with identical swarmalators (adapted from \cite{o2022collective}). (c)--(f) Scatter plots of the $(0,0)$, $(0,S)$, $(S_{1},S_{2})$ where $S_{1}<S_{2}$, and $(S,S)$ states in the $(\theta, x)$ plane. Magenta diamond, green crosses, cyan crosses, and orange stars in panel (a)  show the points where we made the scatter plots for (c)--(f), respectively.}
\label{fig:diagram}
\end{figure}

Numerics shows the system has four steady states which may be categorised by the pair $(S_+, S_-)$. (i) \textit{Async} or (0,0) state: Swarmalators are fully dispersed in space and phase as depicted in Fig.\ref{fig:torus}(a) and Fig.~\ref{fig:diagram}(c). There is no space-phase order so $(S_+, S_-) = (0,0)$. (ii) \textit{Phase waves} or (S,0)/(0,S) state: swarmalators form a band or phase wave \footnote{in 2D this looks like a vortex, see \cite{o2017oscillators}; that's why we called it a 'vortex-like' phase wave in the abstract} where 
$x_i \approx \mp \theta_i$ for $(S,0)$ and $(0,S)$ states, respectively, as depicted in Fig.~\ref{fig:torus}(b) and Fig.~\ref{fig:diagram}(d). In $(\zeta, \eta)$ coordinates, swarmlators are partially locked in $\zeta_i$ and drift in $\eta_i$, or vice versa. 
(iii) Intermediate \textit{mixed} state $(S_{1},S_{2})$ with $S_{1} \neq S_{2} \neq 0$, see Fig.~\ref{fig:diagram}(e): swarmalators form a band along which clusters of correlated swarmalators are moving.(iv) \textit{Sync} or $(S,S)$ state: swarmalators are partially locked in both $\zeta_i$ and $\eta_i$.  
For most initial conditions, two clusters of locked swarmalators separated a distance of $\pi$ in $\zeta, \eta$ 
merge spontaneously, as shown in Fig.\ref{fig:torus}(c) and Fig.~\ref{fig:diagram}(f) (single clusters were also observed.) This `$\pi$-state' results from a symmetry in the model: the transformation $\tilde{x}_i = x_i +\pi$ and $\tilde{\theta}_i= \theta_i +\pi$ leaves Eqs.~(\ref{eq: 1}),(\ref{eq: 2}) unchanged which means a locked swarmalator can be assigned to either cluster without changing the overall dynamic. 
The internal symmetry results in the
formation of mirrored groups of synchronized swarmalators, see \cite{SM}.
Movies of the evolution of these states and demonstrations that they are robust to local coupling (i.e. cutoff beyond a range $\sigma$) are provided in \cite{SM}.


\textit{Generalized OA ansatz.}--- Now we analyze our model by deriving expressions for the order parameters $W_{\pm}$ in each state. Consider the probability $f(v,\omega,\zeta,\eta,t)$ to find a swarmalator with natural velocity $v$, a natural frequency $\omega$, and coordinates $\zeta$ and $\eta$ at time $t$
\begin{equation}
f{\equiv} \frac{1}{N} \sum_{i=1}^{N} \delta(v{-}v_i)\delta(\omega{-}\omega_i)\delta(\zeta{-}\zeta_i)\delta(\eta{-}\eta_i).
\label{eq: 11}
\end{equation}
Differentiating the left and right hand sides of Eq. (\ref{eq: 11}) over $t$ gives the continuity equation,
\begin{eqnarray}
&&\!\!\!\!\!\!\!\!\!\!\frac{\partial f}{\partial t} {+} \frac{\partial}{\partial \zeta} \{ [v{+}\omega - J_{+} S_{+} \sin(\zeta{-}\Phi_{+}) {-} J_{-} S_{-} \sin(\eta {-}\Phi_{-})] f \}
\nonumber \\
&&\!\!\!\!\!\!\!\!\! {+}\frac{\partial}{\partial \eta} \{ [v{-}\omega {-} J_{-} S_{+} \sin(\zeta{-}\Phi_{+}) {-} J_{+} S_{-} \sin(\eta {-}\Phi_{-})] f\}
\nonumber \\
&=&0.
\label{eq: 14}
\end{eqnarray}
\begin{figure}[htpb!]
\centering
\includegraphics[width=8cm,angle=0.]{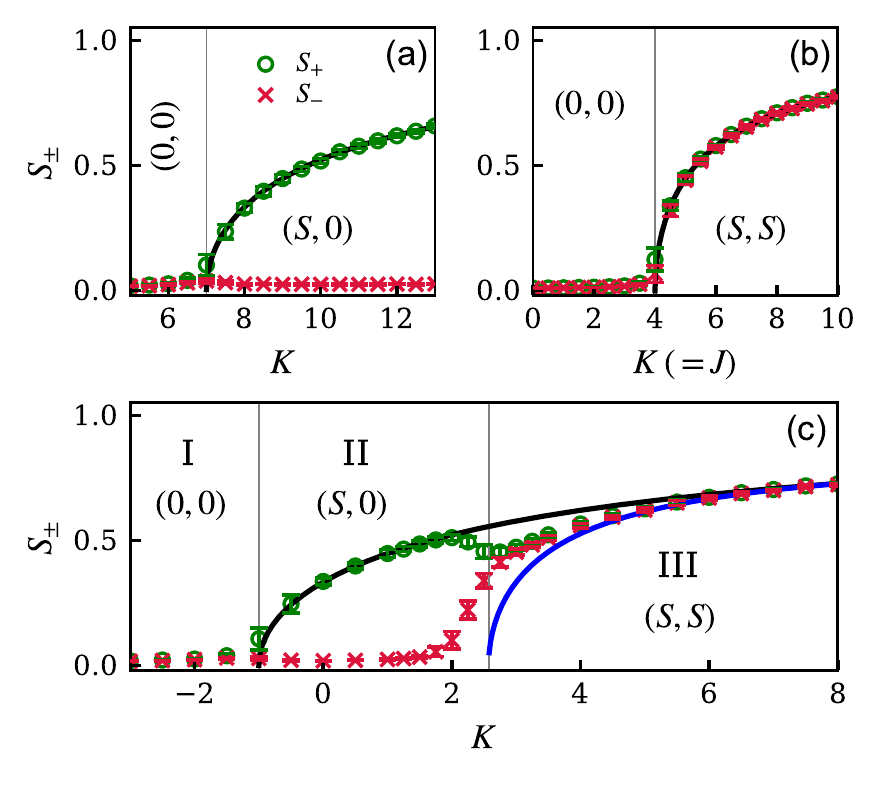}
\caption{Order parameters $S_{\pm}$ versus coupling $K$. (a) Async/phase  wave transition (region II in Fig. \ref{fig:diagram}(a) at $J=1$). (b) Async/sync transition (along the diagonal line in the 
region IV in Fig. \ref{fig:diagram}(a) at $J=K$). (c) 
Async (I)/phase wave (II) /mixed (III)/ sync (IV) transitions at $J=9$. Blue and black solid lines correspond to theoretical expressions Eqs.~(\ref{eq: 28}) and (\ref{eq: 43}), respectively. Black dashed line corresponds to the susceptibility peak, see \cite{SM}. Green open circles and red crosses represent simulation data for the same parameters as in Fig.~\ref{fig:diagram}.}
\label{fig:OP}
\end{figure}
\noindent
Ott and Antonsen showed that for the Kuramoto model, $f$ has an invariant manifold of Poisson kernels (a remarkable finding which effectively solves the model) known as the OA ansatz \cite{ott2008low,ott2009long}. Since our model is a Kuramoto model on the torus, we search for a `torodoidal' OA ansatz: a product of Poisson kernels,
\begin{eqnarray}
&&\!\!\!\!\!\!\!\!\!\!\!\!f(v,\omega,\zeta,\eta,t)=\frac{1}{4 \pi ^2}g_{v}(v)g_{\omega}(\omega)
\nonumber \\
&&\!\!\!\!\!\!\!\!\!\!\!\!\times \Bigl[ 1 {+} \sum_{n =1}^{\infty} \alpha^n e^{i n\zeta} + c.c. \Bigr]\Bigl[ 1 {+} \sum_{m =1}^{\infty} \beta^m e^{i m \eta} + c.c. \Bigr],
\label{eq: 15}
\end{eqnarray}
\noindent
where $\alpha=\alpha(v,\omega,t)$ and $\beta=\beta(v,\omega,t)$ are unknown functions which must be found self-consistently. Substituting Eq.~(\ref{eq: 15}) into Eq.~(\ref{eq: 14}) we find that
$f$ satisfies  Eq. (\ref{eq: 14}) for all harmonics $n$ and $m$ if $\alpha$ and $\beta$ satisfy
\begin{eqnarray}
\frac{d \alpha}{d t} = && - i(v+\omega) \alpha + \frac{1}{2}J_{+} (W^{\ast}_{+}- W_{+} \alpha ^2)
\nonumber \\
&& + \frac{1}{2}J_{-} \alpha (W^{\ast}_{-}\beta^{\ast} - W_{-} \beta),
\label{eq: 16} \\
\frac{d \beta}{d t} = &&- i(v-\omega) \beta +\frac{1}{2}J_{+} (W^{\ast}_{-}- W_{-} \beta ^2)
\nonumber \\
&& + \frac{1}{2}J_{-} \beta (W^{\ast}_{+} \alpha^{\ast}- W_{+} \alpha),
\label{eq: 17}
\end{eqnarray}
in the sub-manifold $\|\alpha\|=\|\beta\|=1$. 
The order parameters $W_{\pm}$ become
\begin{eqnarray}
&&W_{+}=\int_{-\infty}^{\infty} dv \int_{-\infty}^{\infty} d\omega g_{v}(v)g_{\omega}(\omega) \alpha^{\ast}(v,\omega,t),
\label{eq: 18} \\
&&W_{-}=\int_{-\infty}^{\infty} dv \int_{-\infty}^{\infty} d\omega g_{v}(v)g_{\omega}(\omega) \beta^{\ast}(v,\omega,t).
\label{eq: 19}
\end{eqnarray}
Equations (\ref{eq: 16})-(\ref{eq: 19}) comprise a set of self-consistent equations for $W_{\pm}$ in the  $N \rightarrow \infty$ limit. 

\textit{Analysis of async}--- Here swarmalators are uniformly distributed in $x$ and $\theta$ which corresponds to the trivial fixed point $W_{\pm} {=} 0$. Equations~(\ref{eq: 16})-(\ref{eq: 17}) give $\alpha{=}\exp{[i(v{+}\omega)t]}$, $\beta=\exp{[i(v{-}\omega)t]}$. Linearizing around $f = (4\pi)^{-2}$ \cite{SM} reveals the state loses stability at
\begin{equation}
    J_{+,c} = 2 (\Delta_v + \Delta_{\omega})~.
\label{eq: Jc}
\end{equation}
Fig.~\ref{fig:diagram}(a) plots this condition in the $(J,K)$ plane. 

\textit{Analysis of phase waves}--- We analyze the $(S,0)$ phase wave state. 
We look for a solution of Eqs.~(\ref{eq: 16})-(\ref{eq: 19}) that at large time $t$ satisfies: $\dot{\alpha}=0$, $\dot{\beta}\neq0$, $W_{+}\neq 0$ and $W_{-}=0$. We find
\begin{eqnarray}
\alpha(v, \omega)&=& H \Bigl(\frac{v+\omega}{S_{+}J_{+}} \Bigr)~,
\label{eq: 22} \\
\!\!\!\beta(v, \omega,t) &=&\exp \Bigl[-i\frac{JK}{J_{+}} \Bigl(\frac{v}{J}-\frac{\omega}{K} \Bigr) t \Bigr]~,
\label{eq: 23}
\end{eqnarray}
where we introduced a function,
\begin{equation}
H(x) \equiv -ix + \sqrt{1 -x^2 }.
\label{eq: 24}
\end{equation}
Eq.~(\ref{eq: 23}) gives $W_{-}=0$ as desired. Eq.~(\ref{eq: 22}) implies
\begin{equation}
S_{+}= \int_{-\infty}^{\infty} dv \int_{-\infty}^{\infty} d\omega g_{v}(v) g_{\omega}(\omega) H^{\ast}\Bigl(\frac{v+\omega}{S_{+}J_{+}} \Bigr),
\label{eq: 25}
\end{equation}
where we assume $\Phi_+ = 0$ without loss of generality  due to the rotational symmetry. To compute this integral, first observe that if $v$ and $\omega$ are drawn from the Lorentzian distribution, their sum $v+\omega$ is drawn from a Lorentzian with spread $\Delta_v {+}\Delta_{\omega}$. Then integrate over $v+\omega$ using the residue theorem. There is a residue $i(\Delta_v {+} \Delta_{\omega})$ in the upper half complex plane where $H^{\ast}(x)$ is analytic so $S_{+}{=}H^{\ast}[i(\Delta_v {+}\Delta_{\omega})/S_{+}J_{+}]$. Thus,
\begin{equation}
S_{+}= \Bigl[1- \frac{2(\Delta_{v}+\Delta_{\omega})}{J_{+}}\Bigr]^{1/2}.
\label{eq: 28}
\end{equation}
We see $S_+$ bifurcates from $0$ at
\begin{equation}
J_{+,c}=\frac{1}{2}(J+K)= 2(\Delta_{v}+\Delta_{\omega}) 
\label{eq: 29}
\end{equation}
consistent with Eq.~(\ref{eq: Jc}) as the system transitions from the async to the phase wave state (Fig.~\ref{fig:OP}(a)), see the stability analysis in \cite{SM}.
The phase wave $(0,S)$ is a solution of Eqs.~(\ref{eq: 16})-(\ref{eq: 19}) that at large time $t$ satisfies:  $\dot{\alpha}\neq 0$, $\dot{\beta}=0$, $W_{+}=0$, $W_{-}\neq 0$. 

 \textit{Mixed state}--- Here $(S_{1}, S_{2})$, where $S_{1} \neq S_{2}$. This state is intermediate between the phase wave and sync states, see Fig.~\ref{fig:diagram}(a) and compare Fig.~\ref{fig:diagram}(d) and (e). The state with either $S_1 > S_2$ or $S_1 < S_2$ bifurcates from $(S,0)$ or $(0,S)$, respectively. The corresponding order parameters and phase boundaries in $(J,K)$ plane are shown in Figs.~\ref{fig:diagram}(a) and \ref{fig:OP}(c) and discussed in \cite{SM}. The special property of the mixed state is that although $S_1$ and  $S_2$ are time independent, both the functions $\alpha$ and $\beta$ are time dependent in contrast to time independent equations  (\ref{eq: 22}) and (\ref{eq: 33}) (see below) for the phase wave and the sync states. Analytical properties of $\alpha$ and $\beta$ near the boundary with the phase wave are discussed in the Sec. IV, see \cite{SM}.

\textit{Analysis of sync}--- Here $(S_+, S_-) = (S,S)$ so we seek fixed points of Eqs.~(\ref{eq: 16})-(\ref{eq: 17}) with $W_{\pm}{\neq} 0$. We find  
\begin{equation}
\!\!\!\!\!\!\alpha(v, \omega){=}H\Bigl[\frac{v}{JS_{+}}{+}\frac{\omega}{KS_{+}}\Bigr], \,\,\,
\beta(v, \omega) {=} H\Bigl[\frac{v}{JS_{-}}{-}\frac{\omega}{KS_{-}}\Bigr]~.
\label{eq: 33}
\end{equation}
We solve the integrals for $W_{\pm}$ using the residue theorem. This time the natural frequencies combine as $ v/J \pm \omega / K $ which are Lorentzian distributed with spread $\widetilde{\Delta} {\equiv} \Delta_{v}/J{+}\Delta_{\omega}/K $. Equations~$(\ref{eq: 18})$ and $(\ref{eq: 19})$ reduce to $S_{\pm}{=} H^{\ast}(i\widetilde{\Delta}/S_{\pm})$ and so
\begin{equation}
S_{\pm}= \sqrt{1{-} 2\widetilde{\Delta}} \label{eq: 44}
\end{equation}
which bifurcates from $0$ at
\begin{equation}
2\widetilde{\Delta}=2 \Bigl(\frac{\Delta_{v}}{J}+\frac{\Delta_{\omega}}{K} \Bigr)=1.
\label{eq: 43}
\end{equation}
Figure~\ref{fig:diagram}(a) shows this critical curve in the $(J,K)$ plane. Notice it intersects with the critical curve of the phase wave at a point $J=K=2(\Delta_v {+} \Delta_{\omega})$. This means the sync state may bifurcate from the async state directly, without passing through the phase (Fig.~\ref{fig:OP}(b)), which occurs when $J=K$. In this special case, Eqs.~$(\ref{eq: 8})$ and $(\ref{eq: 9})$ for $\dot{\zeta}, \dot{\eta}$ decouple and $W_{\pm}(t)$ may be solved for all $t$ (see \cite{SM}). In the generic case $J \neq K$, however, the sync state bifurcates from the intermediate mixed state (Fig.~\ref{fig:diagram}(c)). As is evident from Fig.~\ref{fig:OP}(c), the point $J=K=2(\Delta_v {+} \Delta_{\omega})$ is a tetracritical point, at which four phases (async, sync, phase wave, and mixed) meet. The appearance of the sync state can be considered as the separation of dense clusters of locked swarmalators with time-independent coordinates and dilute drifting swarmalators in $(x, \theta)$ space. This phenomenon is qualitatively similar to motility induced phase separation observed in self-propelled particles and various microorganisms, see for example \cite{cates2015motility}.


To back up these numerical tests of our results we performed four additional analyses. First, we re-derive $S_{\pm}$ using a microscopic, swarmalator-level, approach (as opposed to the macroscopic, density-level approach the OA ansatz is based on). In the phase wave $(S,0)$, swarmalators are partially locked in $\dot{\zeta}_i = 0$ and drift in $\dot{\eta}_i \neq 0 $. Applying these conditions to Eqs.~(\ref{eq: 8}) and (\ref{eq: 9}) yields
\begin{align}
& \sin(\zeta_i {-} \Phi_{+}) = \frac{v_i +\omega_i}{S_{+}J_{+}} \label{eq: 33b} \\
& \eta_i(t) = \eta_i(0) + \frac{1}{J_{+}}(Kv_i -J\omega_i)t ~, \label{eq: 34b}
\end{align}
\noindent
where $-S_{+}J_{+} \leq v_i +\omega_i \leq S_{+}J_{+}$ and $\eta_i(0)$ is an initial phase. Following Kuramoto \cite{kuramoto2003chemical}, the order parameter must be self-consistent: $S_+ := N^{-1} \sum_j e^{i \zeta_j}$. Plugging Eq.~(\ref{eq: 33b}) indeed gives the expression Eq.~(\ref{eq: 28}) for $S_{+}$ in agreement with the generalized OA ansatz (Similarly, Eq.~(\ref{eq: 34b}) implies $S_- = 0$ as expected). We also attempted a microscopic analysis of the sync state but the calculations were beyond the scope of this Letter \cite{SM}. Second, we checked the identical swarmalator limit which has been analyzed previously (without an OA ansatz) \cite{o2022collective}. As $\Delta_v, \Delta_{\omega} \rightarrow 0,0$, the critical curve for the phase wave Eq.~(\ref{eq: 29}) approaches $J+K=0$, while that of the sync state Eq.~(\ref{eq: 43}) approaches $J,K>0,0$ in agreement with \cite{o2022collective}. Fig.~\ref{fig:diagram}(b) plots these in $(J,K)$ space to allow a visual comparison. Third, we calculated the stability of async using the OA equations (Eqs.~(10)-(11)) and found it agreed with Eq.~\eqref{eq: Jc} \cite{SM} (derived by perturbing the continuity equation \cite{SM}). Fourth, we used the OA equations to derive $S_{sync}$ and its critical coupling $K_c$ for a simpler distribution $g_{\omega(v)}(x) = \frac{1}{2}\delta(x-\Delta) + \frac{1}{2} \delta(x + \Delta)$ which agreed with simulation perfectly \cite{SM}. This completes our analysis.

{\textit{`Hidden' phase transition}---We close by pointing out a curious feature of the swarmalator model. At $J = 0$, the positions evolves at constant speed $\dot{x_i} = v_i \Rightarrow x_i = v_i t$ which means the phases obey 
\begin{equation}
\dot{\theta}_i=\omega_i+\frac{K}{N} \sum_{j=1}^N \sin(\theta_j - \theta_i) \cos[(v_j-v_i)t].
\label{eq: 44}
\end{equation}
One can think of this equation as a model for a group of oscillators with random, time-dependent couplings. In turn, the results presented in this letter reveal a phase transition hidden in the time-dependence of $\theta_i$, which extends to the case where $J=0$. This `hidden' phase transition causes incoherent oscillators to become phase-locked at $\theta_i = - v_i t + \zeta_i /2$ (the $(S, 0)$ state) or $\theta_i = v_i t - \eta_i /2$ (the $(0, S)$) where $\zeta_i$ $(\eta_i)$ is the phase from Eq.~(\ref{eq: 44}). Curiously, if we reinterpret $v_i t$ as a heterogeneous field acting on the couplings, we see that the oscillators have become tuned to the field frequency $v_i$. To the best of our knowledge, this is a novel result and may provide a useful means for tuning a population of oscillators to a prescribed set of frequencies in an experimental setting.

To conclude, we have presented a simple, solvable model of 
swarmalators. The model has a rich phase diagram with 
a \textit{tetracritical} point at which four phases meet.
The model also captures the behavior of 
real-world swarmalators/active matter such as groups of sperm \cite{creppy2016symmetry} and vinegar eels \cite{quillen2021synchronized,quillen2021metachronal} (which swarm in quasi-1D rings), and the rotational component of 2D, real-world  swarmalators such as forced colloids \cite{zhang2020reconfigurable,yan2012linking,yan2015rotating}. Our simulations showed that the cutoff in the spatial interaction kernel does not qualitatively change the dynamics of swarmalators in comparison to global coupling \cite{SM}. Thus, the exact solution of the swarmalator model with all-to-all coupling should have applicability to a variety of situations with local coupling. We hope our work will be useful to the active matter community, as it provides a new toy model, and interesting to the sync community, as the first  OA ansatz for oscillators which are mobile (mobile in a 1D periodic domain, at least).

Future work could study the stability of the phase wave, mixed, and sync states (note we derived criteria for their \textit{existence} only). 
Incorporating delayed interactions or external forcing -- which are 
analyzable with our OA ansatz -- would also be interesting. Finally, our model and predictions could be experimentally tested in circularly confined colloids or robotic swarms \cite{barcis2020sandsbots,barcis2019robots}.

This work is funded by national funds (OE) through Portugal's FCT Funda\c{c}\~{a}o para a Ci\^{e}ncia e Tecnologia, I.P., within the scope of the framework contract foreseen in paragraphs 4,5 and 6 of article 23, of Decree-Law 57/2016, of August 29, and amended by Law 57/2017, of July 19. Code used in simulations available at \footnote{https://github.com/Khev/swarmalators/tree/master/1D/on-ring/non-identical}.

    

%

\pagebreak
\widetext
\begin{center}
\textbf{\large Supplemental Materials: sync and swarm: solvable model of nonidentical swarmalators}
\end{center}
\setcounter{equation}{0}
\setcounter{figure}{0}
\setcounter{table}{0}
\setcounter{page}{1}
\makeatletter
\renewcommand{\theequation}{S\arabic{equation}}
\renewcommand{\thefigure}{S\arabic{figure}}
\renewcommand{\bibnumfmt}[1]{[S#1]}
\renewcommand{\citenumfont}[1]{S#1}

\section{Stability of async state within the microscopic approach}
In the $N \rightarrow \infty$ limit the async state is given by the density $\rho(\zeta,\eta, \omega_+, \omega_-, t) = (4 \pi)^{-2}$. Perturbing around this state is density space give us the critical coupling strengths. The density obeys the continuity equation
\begin{align}
    \dot{\rho} + \nabla \cdot ( v \rho ) = 0 \\
    \dot{\rho} + v \cdot \nabla \rho + \rho \nabla \cdot v = 0 
    \label{cont1}
\end{align}
where the velocity $v$ is interpreted in the Eulerian sense and is given by the right hand side of the model equations, listed here again for convenience,
\begin{align}
\dot{\zeta}_i &=  v_i {+} \omega_i {-} J_{+}S_{+} \sin(\zeta_i {-} \Phi_{+}) {-} J_{-}S_{-} \sin(\eta_i {-} \Phi_{-}),
\label{xi_eom} \\
\dot{\eta}_i &= v_i {-} \omega_i {-} J_{-} S_{+} \sin(\zeta_i {-} \Phi_{+}) {-} J_{+}S_{-} \sin(\eta_i {-} \Phi_{-}).
\label{eta_eom}
\end{align}
Consider the perturbation
\begin{equation}
    \rho = \rho_0 + \epsilon \rho_1 = (4 \pi^2)^{-1} + \epsilon \rho_1(\zeta, \eta, t) 
    \label{peturb1}
\end{equation}
Normalization requires $\int \rho(x,\theta) = 1$ which implies 
\begin{equation}
    \int \rho_1(x,\theta,t) dx d \theta = 0
    \label{norm}
\end{equation}
The density ansatz \eqref{peturb1} decomposes the velocity
\begin{equation}
    v = v_0 + \epsilon v_1 
    \label{vel}
\end{equation}
\noindent
where $v_{0} = (v + \omega, v - \omega)$ is the velocity in the async state. The perturbed velocity $v_1$ is given by Eqs. \eqref{xi_eom}, \eqref{eta_eom} with the order parameters  perturbed: 
\begin{align}
    W_+^{(1)} &= \int e^{i \zeta} \rho_1(\zeta, \eta, t) d\zeta d \eta \\
    W_-^{(1)} &= \int e^{i \eta} \rho_1(\zeta, \eta, t) d \zeta d \eta
\end{align}
Plugging the perturbation Eq.~\eqref{peturb1} into the continuity equation \eqref{cont1} yields
\begin{equation}
    \dot{\rho_1} + v_0 \nabla \cdot \rho_1 + \rho_0 \nabla \cdot v_1 = 0 \label{t4}
\end{equation}
Let's tackle the divergence term first. Writing $v_1$ in complex exponentials,
\begin{align}
    v_{1,\zeta} &=  \frac{J_+}{2 i}( W_+^{(1)} e^{- i\zeta} - \bar{W_+}^{(1)}e^{i \zeta}) + \frac{J_-}{2 i}( W_-^{(1)} e^{- i\eta} - \bar{W_-}^{(1)}e^{i \eta}) \\
     v_{1,\eta} &= \frac{J_-}{2 i}( W_+^{(1)} e^{- i \zeta} - \bar{W_+}^{(1)}e^{i \zeta}) + \frac{J_+}{2 i}( W_-^{(1)} e^{- i\eta} - \bar{W_-}^{(1)}e^{i \eta})    \label{divv1}
\end{align}
Then
\begin{align}
    \nabla \cdot v_1 &= -\frac{J_+}{2}( W_+^{(1)} e^{-i \zeta} + 
    \overline{W}_{-}^{(1)} e^{i \zeta} ) - \frac{J_+}{2}( W_-^{(1)} e^{- i\eta} - 
    \overline{W}_{-}^{(1)}e^{i \eta}) 
\end{align}

Next we expand $\rho_1$ in a Fourier series,
\begin{align}
    \rho_1(\zeta, \eta, v, \omega, t) &= \frac{1}{4 \pi^2} \Big( 1 + c_+(v, \omega, t) e^{-i \zeta} + c_-(v, \omega, t) e^{-i \eta} + c.c.
    + \rho_1^{\dagger}(\zeta, \eta, v, \omega, t) \Big). \label{rho1_fs}
\end{align}
where $\rho_1^{\dagger}$ contains all the higher harmonics (we abuse notation slightly by dropping the subscripts in the $g_{\omega}(\omega)$ and $g_{v}(v)$). Note this implies $W_{\pm}^{(1)} = \int c_{\pm}(v, \omega, t) g(\omega) g(v) d \omega d v$. Plugging Eq.~\eqref{rho1_fs} and the expression for the divergence Eq.~\eqref{divv1} into Eq.~\eqref{t4} and projecting onto the $e^{i\zeta}, e^{i \eta}$ modes leads to 
\begin{align}
    \dot{c}_{\pm}= i (v \pm \omega) c_{\pm} - \frac{J_+}{2} \int c_{\pm}(v, \omega) g(v) g(\omega) d v d \omega \label{t5}
\end{align}

Notice each $c_{\pm}$ obeys the same stability equation as the regular Kuramoto model \cite{strogatz1991stability}. This stability properties are shown in \cite{strogatz1991stability} to be interesting (the incoherent state turns out to be linearly neutrally stable). Here we repeat their analysis. \\

We seek the discrete spectrum $c_{\pm} = b_{\pm}(v, \omega, t) e^{i \lambda_{\pm}t }$. Subbing this in yields,
\begin{align}
    \lambda_{\pm} b_{\pm} &= i (v \pm \omega) b_{\pm}  - \frac{J_+}{2} \int b_{\pm}(v, \omega) g(v) g(\omega) d v d \omega \\
    \lambda_{\pm} b_{\pm} &= i (v \pm \omega) b_{\pm}  - \frac{J_+}{2} A \\
    b_{\pm} &= \frac{J_+}{2} \frac{A}{\lambda + i (v \pm \omega)} 
\end{align}

where, crucially, $A$ is a constant. Self-consistency requires
\begin{align}
    A &= \int b_{\pm}(v, \omega) g(v) g(\omega) d v d \omega  \\
    A &= \int \frac{J_+}{2} \frac{A}{\lambda + i (v \pm \omega)} g(v) g(\omega) d v d \omega  \\
    1 &= \frac{J_+}{2} \int \frac{g(v) g(\omega) }{\lambda + i (v \pm \omega)} d v d \omega 
\end{align}
It can be shown that there is precisely one, real, solution to the above equation in which case we rewrite
\begin{align}
    1 &= \frac{J_+}{2} \int \frac{ \lambda}{\lambda + (v \pm \omega)^2} g(v) g(\omega)  d v d \omega
\end{align}
The interesting feature here is that $\lambda < 0$ never exists (see \cite{strogatz1991stability} for a discussion about this; it is related to Landau dampling). Regardless, for Lorentzian $g(\omega/v)$ this may be evaluated exactly,
\begin{align}
    1 &= \frac{J_+}{2} \Big( \frac{1}{ \Delta_v + \Delta_{\omega} + \lambda} \Big) \\
    \lambda_{\pm} &= \frac{1}{2}( J_+ -2(\Delta_v  + \Delta_{\omega} ) )
\end{align}
Setting $\lambda = 0$ yields

\begin{align}
J_{+,c} = 2(\Delta_v + \Delta_{\omega})
\end{align}

in agreement with the main text. Recall this expression for $\lambda$ was for the discrete spectrum only. It can be shown the continuous spectrum lies on the imaginary axis $\lambda_{\pm} = i(v \pm \omega)$ (see \cite{strogatz1991stability}). \\


\section{Stability of the async state within the generalized OA ansatz}
In the main text, we introduced the generalized OA ansatz and derived the set of equations Eqs. (10)--(13) (listed below) that must be solved self-consistently
\begin{eqnarray}
\frac{d \alpha}{d t} = && - i(v+\omega) \alpha + \frac{1}{2}J_{+} (W^{\ast}_{+}- W_{+} \alpha ^2)
+ \frac{1}{2}J_{-} \frac{\alpha}{\beta} (W^{\ast}_{-}- W_{-} \beta ^2),
\label{eq: 16_SM} \\
\frac{d \beta}{d t} = &&- i(v-\omega) \beta +\frac{1}{2}J_{+} (W^{\ast}_{-}- W_{-} \beta ^2)
+ \frac{1}{2}J_{-} \frac{\beta}{\alpha} (W^{\ast}_{+}- W_{+} \alpha ^2).
\label{eq: 17_SM}
\end{eqnarray}
\begin{eqnarray}
&&W_{+}=\int_{-\infty}^{\infty} dv \int_{-\infty}^{\infty} d\omega g_{v}(v)g_{\omega}(\omega) \alpha^{\ast}(v,\omega,t),
\label{eq: 18_SM} \\
&&W_{-}=\int_{-\infty}^{\infty} dv \int_{-\infty}^{\infty} d\omega g_{v}(v)g_{\omega}(\omega) \beta^{\ast}(v,\omega,t).
\label{eq: 19_SM}
\end{eqnarray}
In the async state, the order parameters $W_{\pm}$ are zero. Let us determine the region of stability of the async state to small perturbations. First we find the functions $\alpha$ and $ \beta$ in the async state. At $W_{\pm}=0$, the equations (\ref{eq: 16_SM}) and (\ref{eq: 17_SM}) have a solution without loss of generality,
\begin{eqnarray}
\alpha_{0}(v,\omega,t)=\exp[-i(v+\omega)t],
\label{eq: async1_SM} \\
\beta_{0}(v,\omega,t)=\exp[-i(v-\omega)t)].
\label{eq: async2_SM}
\end{eqnarray}
In order to check the self-consistency of this solution, we substitute these functions into Eqs. (\ref{eq: 18_SM}) and (\ref{eq: 19_SM}). We obtain $W_{\pm}=\exp[-(\Delta_v +\Delta_{\omega})t)]\rightarrow 0$ in the limit $t \rightarrow \infty$. Thus, the solution given by Eqs. (\ref{eq: async1_SM}) and (\ref{eq: async2_SM}) is self-consistent.
Let us consider the stability of the async state. For this purpose, we consider a small perturbation about the solution given by Eqs. (\ref{eq: async1_SM}) and (\ref{eq: async2_SM}):
\begin{eqnarray}
\alpha(v,\omega,t)=\alpha_{0}(v,\omega,t)+\alpha_{1}(v,\omega,t),
\label{eq: pert1_SM} \\
\beta(v,\omega,t)=\beta_{0}(v,\omega,t)+\beta_{1}(v,\omega,t),
\label{eq: pert2_SM} \\
W_{+}^{(1)}=\int_{-\infty}^{\infty} dv \int_{-\infty}^{\infty} d\omega g_{v}(v)g_{\omega}(\omega) \alpha^{\ast}_{1}(v,\omega,t),
\label{eq: pert3_SM} \\
W_{-}^{(1)}=\int_{-\infty}^{\infty} dv \int_{-\infty}^{\infty} d\omega g_{v}(v)g_{\omega}(\omega) \beta^{\ast}_{1}(v,\omega,t).
\label{eq: pert4_SM}
\end{eqnarray}
We assume that $\| \alpha_{1}(v,\omega,t)\|, \|\beta_{1}(v,\omega,t)\|, \|W_{\pm}^{(1)}\|\ll 1$. In the first order, the Eqs. (\ref{eq: 16_SM}) and (\ref{eq: 17_SM}) take a form
\begin{eqnarray}
\frac{d \alpha_1}{d t} = - i(v+\omega) \alpha_1 + \frac{1}{2}J_{+} W^{(1) \ast}_{+}- \frac{1}{2}J_{+} W^{(1)}_{+} \alpha_{0}^2
+ \frac{1}{2}J_{-} \frac{\alpha_{0}}{\beta_{0}} (W^{(1)\ast}_{-}- W^{(1)}_{-} \beta_{0}^2),
\label{eq: pert5_SM} \\
\frac{d \beta_1}{d t} = - i(v-\omega) \beta_1 +\frac{1}{2}J_{+} W^{(1)\ast}_{-}- \frac{1}{2}J_{+} W^{(1)}_{-} \beta_0 ^2
+ \frac{1}{2}J_{-} \frac{\beta_0}{\alpha_0} (W^{(1)\ast}_{+}- W^{(1)}_{+} \alpha_{0} ^2).
\label{eq: pert6_SM}
\end{eqnarray}
Note that at $t \gg 1$  the terms with $\alpha_0$ and $\beta_0$ are rapidly oscillating functions of $v$ and $\omega$. When integrating over $v$ and $\omega$, a contribution of these rapidly oscillating terms is negligibly small (this is a  common argument in physics for approximating integrals). The function $\alpha(v,\omega,t)$ becomes a function of a variable $y \equiv v+ \omega$, i.e., $\alpha(v,\omega,t)=\alpha(y,t)$. Then the integration of the left and right hand sides of Eqs. (\ref{eq: pert5_SM}) and (\ref{eq: pert6_SM}) over $v$ and $\omega$ gives
\begin{eqnarray}
\frac{d W_{+}^{(1)}}{d t}=\int_{-\infty}^{\infty} dv \int_{-\infty}^{\infty} d\omega g_{v}(v)g_{\omega}(\omega) \frac{d \alpha^{\ast}_1}{d t}= \int_{-\infty}^{\infty} dy G(y)\frac{d \alpha^{\ast}_1(y)}{d t}= - (\Delta_v +\Delta_{\omega} - \frac{1}{2}J_{+}) W_{+}^{(1)} ,
\label{eq: pert7_SM} \\
\frac{d W_{-}^{(1)}}{d t}=\int_{-\infty}^{\infty} dv \int_{-\infty}^{\infty} d\omega g_{v}(v)g_{\omega}(\omega) \frac{d \beta^{\ast}_1}{d t}= \int_{-\infty}^{\infty} dy G(y)\frac{d \beta^{\ast}_1(y)}{d t}=- (\Delta_v +\Delta_{\omega} - \frac{1}{2}J_{+}) W_{-}^{(1)},
\label{eq: pert8_SM}
\end{eqnarray}
where we introduced a function $G(y)$,
\begin{equation}
G(y)  \equiv \int_{-\infty}^{\infty} dv \int_{-\infty}^{\infty} d\omega g_{v}(v) g_{\omega}(\omega) \delta[y-(v \pm \omega)]=\frac{\Delta_{v}+\Delta_{\omega}}{\pi[y^2+ (\Delta_{v}+\Delta_{\omega})^2]},
\label{eq: 1s11a}
\end{equation}
and used the fact that it has a residue $y=i(\Delta_v +\Delta_{\omega})$ in the upper half plane where $\alpha^{\ast}_0(y)$ and $\beta^{\ast}_0(y)$ are analytical. Equations (\ref{eq: pert7_SM}) and (\ref{eq: pert8_SM}) show that the async state is stable if $\Delta_v +\Delta_{\omega} - \frac{1}{2}J_{+} >0$. The perturbation decreases with increasing time and $W_{\pm}^{(1)}(t) \rightarrow 0$ at $t \rightarrow \infty$. The critical line in $(J,K)$ plane is $J_{+,c}= 2(\Delta_v +\Delta_{\omega})$ that agrees with Eq. (20) in the main text, see Fig.~2. Above the critical line the swarmalators form the phase wave states at $J \neq K$.

\section{Stability of the phase wave state at $J=0$ and $K\neq0$}
In this section we consider the stability of the phase wave states $(S,0)$ and $(0,S)$ for the couplings $J=0$ and $K\neq0$. We discussed this case in the main text in the context of the `hidden'  phase transition. At $J=0$ we have $J_{+}=K/2$ and  $J_{-}= - K/2$. Using Eqs. (\ref{eq: 16_SM}) and (\ref{eq: 17_SM}) and summing $\beta \dot{\alpha} + \alpha \dot{\beta}$, we find
\begin{equation}
\frac{d (\alpha \beta) }{d t} = - 2 i v \alpha \beta.
\label{eq: pert9_SM}
\end{equation}
Therefore, 
\begin{equation}
\alpha \beta = e^{- 2 i v t}.
\label{eq: pert10_SM}
\end{equation}
Substitution of $ \beta =  \exp(- 2 i v t)/\alpha $ into Eqs. (\ref{eq: 16_SM}), (\ref{eq: 18_SM}), and (\ref{eq: 19_SM}) gives a set of self-consistent equations for the function $ \alpha $ and the order parameters $W_{\pm}$,
\begin{eqnarray}
\frac{d \alpha }{d t} &=& - i(v + \omega) \alpha + \frac{K}{4}(W^{\ast}_{+}- W_{+} \alpha ^2)
- \frac{K}{4} \Bigl[ \alpha^2 W^{\ast}_{-}e^{2 i v t}-  W_{-} e^{-2 i v t} \Bigr].
\label{eq: pert11_SM} \\
W_{+}&=&\int_{-\infty}^{\infty} dv \int_{-\infty}^{\infty} d\omega g_{v}(v)g_{\omega}(\omega) \alpha^{\ast}(v,\omega,t),
\label{eq: 18b_SM} \\
W_{-}&=&\int_{-\infty}^{\infty} dv \int_{-\infty}^{\infty} d\omega g_{v}(v)g_{\omega}(\omega)\frac{e^{2 i v t}}{\alpha^{\ast}} .
\label{eq: 19b_SM}
\end{eqnarray}
This set of equations has two solutions, corresponding to the $(S,0)$ and $(0,S)$ phase wave states, respectively. To find a solution for the $(S,0)$ state,  in the limit $t \gg 1$ we neglect  all terms with rapidly oscillating functions  $\exp(\pm 2 i v t)$ with respect to $v$. Integrating over $v$ and $\omega$ in Eqs. (\ref{eq: pert11_SM}) and (\ref{eq: 19_SM}), we find that the contribution of the terms containing the rapidly oscillating functions $\exp(\pm 2 i v t)$ decreases exponentially in time and tends to zero  at $t \rightarrow \infty$, therefore $W_{-} \rightarrow 0$. The equation (\ref{eq: pert11_SM}) is reduced to the evolution equation for the regular Kuramoto model. Therefore, at $J=0$ the phase wave state $(S,0)$, see  Eq. (15) in the main text, is stable above the critical point $K_c = 4 (\Delta_v + \Delta_{\omega})$ in agreement with Eq. (20) in the main text. In order to find a solution for the $(0,S)$ state, we use a substitution $\alpha= \exp(2 i v t)/\beta $ where $\beta$ does not oscillate in the limit $t \gg 1$. In this limit the function $\beta$ relaxes to a solution of Eqs. (\ref{eq: 16_SM}) and (\ref{eq: 17_SM}) with $\dot{\alpha}\neq 0$, $\dot{\beta}=0$, $W_{+}=0$, and $W_{-}=S_{-}\neq 0$. The set of equations has also an unstable fixed point corresponding to state $(0,0)$ and described by Eqs. (\ref{eq: async1_SM})--(\ref{eq: async2_SM}). Any perturbation about this point leads either to $(S,0)$ or to $(0,S)$ state. It is similar to the Ising model below critical temperature where there is an unstable paramagnetic fixed point between two ordered states with spins up or down. \\

 \section{Emergence of the mixed $(S_1, S_2)$ state } 
The mixed state is a combination of the phase wave state, in which the locked swarmalators execute shear flow in the $\hat{\xi}$ or  $\hat{\eta}$ direction, and the sync state, in which the locked swarmalators sit at fixed points. (Recall in both states there are drifting swarmalators which continually move in the $(\hat{\xi}, \hat{\eta})$ directions). We picture this hybrid behavior as a river (the phase wave) with chunks of ice (the synced swarmalators) quivering on its surface. It's best viewed in Supplementary Movie 5. \\ 

Here we analyze this mixed $(S_1, S_2)$ state by studying the destabilization of the phase wave $(S,0)$ state from which it bifurcates. We assuming without loss of generality that $S_1 < S_2$; the symmetric $(S_2, S_1)$ state is also realized and bifurcates from the $(0,S)$ state. We consider the general case $J \neq K$ and derive the critical line shown in Figs. 2(a) and 3(c) in the main text.

In the sub-manifold $|\alpha|=|\beta|=1$, we look for the functions $\alpha$ and $\beta$ in form
\begin{equation}
\alpha=e^{ia},  \,\,\,\, \beta=e^{ib},
\label{eq: pws0}
\end{equation}
where the phases $a=a(v,\omega,t)$ and $b=b(v,\omega,t)$ are real functions. Taking into account the rotational symmetry, in this representation the generalized OA equations and the order parameters (10)-(13) of the main text take a form
\begin{eqnarray}
\dot{a} &=&-v-\omega -J_+ S_+ \sin a-J_- S_- \sin b,
\label{eq: pws1} \\
\dot{b} &=&-v+\omega -J_+ S_- \sin b-J_- S_+ \sin a,
\label{eq: pws2} \\
S_+ &=& \int_{-\infty}^{-\infty} dv \int_{-\infty}^{-\infty} d\omega g_{v}(v)g_{\omega}(\omega) \cos a,
\label{eq: pws3} \\
S_- &=& \int_{-\infty}^{-\infty} dv \int_{-\infty}^{-\infty} d\omega g_{v}(v)g_{\omega}(\omega) \cos b.
\label{eq: pws4}
\end{eqnarray}
The phase wave state $(S,0)$ is determined by the following conditions: $\dot{a}=0$, $\dot{b} \neq 0$, $S_+ \neq 0$ and $S_- =0$. In this case, Eqs. ~(\ref{eq: pws1})-(\ref{eq: pws4}) give
\begin{eqnarray}
\sin a_0 &=&-\frac{v+\omega}{J_+ S_{+}^{(0)}},
\label{eq: pws5} \\
b_0 &=&- rt,
\label{eq: pws6} \\
S_{+}^{(0)}&=&\sqrt{1-2(\Delta_v +\Delta_{\omega})/J_+}~,
\label{eq: pws7}
\end{eqnarray}
where we define
\begin{equation}
r \equiv v+\omega +J_- S_{+}^{(0)} \sin a_0= - \frac{JK}{J_+} \Bigl(\frac{v}{J}-\frac{\omega}{K} \Bigr).
\label{eq: pws8}
\end{equation}
The lower and upper subscript `0' means that these functions describe the unperturbed state $(S,0)$ with the order parameter $S=S_{+}^{(0)}$ from Eq. (19) of the main text. \\

We aim to find a region of parameters $J$ and $K$ when the state $(S_+,S_-)$ with  time-independent order parameters $S_{\pm}$ appears from $(S,0)$ state. For this purpose we assume that if $S_{-}^{(1)} \equiv S_{-} \ll 1$, then changes of the phases $a$ and $b$ and the the order parameter $S_+$ with respect to $a_0$, $b_0$, and $S_{+}^{(0)}$ are also small,
\begin{eqnarray}
a&=&a_0 + a_1,
\label{eq: pws9} \\
b&=&b_0 + b_1,
\label{eq: pws10} \\
S_+ &=& S_0 + S_{0}^{(1)},
\label{eq: pws11}
\end{eqnarray}
where $|a_1|\ll |a_0|$, $|b_1|\ll |b_0|$, and $S_{+}^{(1)} \ll 1$. In order to find $a_1(t)$, $b_1 (t)$, and $S_{\pm}^{(1)}$, we solve Eqs.~(\ref{eq: pws1})-(\ref{eq: pws4}) in the first order of the perturbation theory:
\begin{eqnarray}
\dot{a_1} &=& -J_+ S_{+}^{(1)} \sin a_0 -J_+ S_{+}^{(0)} a_1 \cos a_0 - J_- S_{-}^{(1)} \sin b_0,
\label{eq: pws12} \\
\dot{b_1} &=&-J_- S_{+}^{(1)} \sin a_0 -J_- S_{+}^{(0)} a_1 \cos a_0 -  J_+ S_{+}^{(1)} \sin b_0,
\label{eq: pws13} \\
S_{+}^{(1)} &=& - \int_{-\infty}^{-\infty} dv \int_{-\infty}^{-\infty} d\omega g_{v}(v)g_{\omega}(\omega) a_1 \sin a_0 ,
\label{eq: pws14} \\
S_{-}^{(1)} &=& - \int_{-\infty}^{-\infty} dv \int_{-\infty}^{-\infty} d\omega g_{v}(v)g_{\omega}(\omega)  b_1 \sin b_0.
\label{eq: pws15}
\end{eqnarray}
We obtain
\begin{eqnarray}
\!\!\!a_1(t) &{=}& - \frac{1}{\gamma}J_+ S_{+}^{(1)} \sin a_0 (1- e^{-\gamma t}) -\frac{J_- S_{-}^{(1)}}{\gamma^2 +r^2} [r \cos (rt) - \gamma \sin (rt)] + \frac{J_- S_{-}^{(1)}r  e^{-\gamma t}}{\gamma^2 +r^2},
\label{eq: pws16} \\
\!\!\!b_1 (t)&{=}& \frac{J_{-}^2 S_{-}^{(1)}S_{+}^{(0)} \cos  a_0 }{\gamma^2 +r^2} [\sin (rt) {+} \frac{\gamma}{r} (\cos (rt){-}1)] {-} \frac{J_+ S_{-}^{(1)}}{r}(\cos (rt){-}1) {-} \frac{J_{-}^2 S_{-}^{(1)}S_{+}^{(0)} r \cos a_0 }{\gamma(\gamma^2 +r^2)} 
\nonumber \\
&-&\frac{1}{\gamma}J_{-}S_{+}^{(1)} \sin a_0,
\label{eq: pws17}
\end{eqnarray}
where we define $\gamma= J_+ S_{+}^{(0)} \cos a_0$. Substituting these functions into Eqs.~(\ref{eq: pws14}) and (\ref{eq: pws15}),  we obtain a set of equations for $S_{\pm}^{(1)}$ that in the limit $t \gg 1$  take a form,
\begin{eqnarray}
S_{+}^{(1)} &=&  A_+(J,K) S_{+}^{(1)},
\label{eq: pws18} \\
S_{-}^{(1)} &=& A_-(J,K) S_{-}^{(1)},
\label{eq: pws19}
\end{eqnarray}
where the coefficients $A_{\pm}(J,K)$ equal to 
\begin{eqnarray}
A_{+}(J,K)&=&  J_+ \int_{-\infty}^{\infty} dv \int_{-\infty}^{\infty}d\omega g_{v}(v)g_{\omega}(\omega)\frac{\sin^{2}a_0}{\gamma}\Theta\Bigl[(J_+S_{+}^{(0)})^2 - (v+  \omega)^2 \Bigr]=\frac{\Delta_v+\Delta_{\omega}}{2J_+ - \Delta_v - \Delta_{\omega}},
\label{eq: pws20} \\
A_{-}(J,K) &=& \frac{1}{2}J_-^2S_+^{(0)}\int_{-\infty}^{\infty} dv \int_{-\infty}^{\infty}d\omega g_{v}(v)g_{\omega}(\omega)\frac{\cos a_0}{\gamma^2+r^2}\Theta\Bigl[(J_+S_{+}^{(0)})^2 - (v+  \omega)^2 \Bigr] + \frac{1}{2}\Bigr[ \frac{\Delta_v}{K} + \frac{\Delta_\omega}{J}\Bigl]^{-1}.
\label{eq: pws21}
\end{eqnarray}
The last term in Eq.~(\ref{eq: pws21}) was obtained by use of the equality
\begin{equation}
\lim_{t\rightarrow \infty} \int_{-\infty}^{\infty} dr F(r) \frac{\sin rt}{r}=\pi F(0).
\label{eq: pws22}
\end{equation}
where $F(r)$ is an analytical function. According to Eq.~(\ref{eq: pws20}), the coefficient $A_+(J,K)$ decreases from 1 to 0 when increasing $J_+= (J+K)/2$ from the critical value  $J_{+,c}=2(\Delta_v+\Delta_{\omega})$ to $\infty$. It means that in the first order in $S_{-}^{(1)}$, the Eq.~(\ref{eq: pws18}) has the only one solution $S_{+}^{(1)}=0$ at $J_+ > J_{+,c}$. With increasing the couplings the coefficient $A_-(J,K)$ increases from zero. 
In the space $(J,K)$,  a critical line of the phase transition from the phase wave state $(S_+,0)$ into the intermediate mixed state $(S_+,S_-)$ with non-zero order parameter $S_-$ is determined by an equation
\begin{equation}
A_-(J,K)=1~.
\label{eq: pws23}
\end{equation}
This critical line is shown in Fig.~2(a) of the main text.

\section{Susceptibility near a transition into $(S,0)/(0,S)$, $(S_{1},S_{2})$, and $(S,S)$ states} 
In order to confirm our analytical calculations of the critical line Eq.~(\ref{eq: pws23}) of the phase transition into the intermediate mixed state $(S_{1},S_{2})$ and to understand the unusual behavior of the swarmalator model near the transition into the sync state $(S,S)$ in Fig. 3(c) of the main text, we carried out numerical simulations of the swarmalator model and found so-called `susceptibility' as a function of the couplings $J$ an $K$. The susceptibility was introduced for the Kuramoto model in \cite{yoon2015critical}. The importance of the susceptibility is that it has a peak (divergence in the infinite size limit) at a critical point of the synchronization transition. The susceptibility $\chi_{\pm}$ corresponding to  the complex order parameters $W_{\pm}$ is defined as follows,
\begin{equation}
\chi_{\pm}=N\Bigl[\langle W_{\pm}(t)W_{\pm}^{*}(t)\rangle_t - \langle W_{\pm}(t)\rangle_t \langle W_{\pm}^{*}(t)\rangle_t\Bigr]
\label{sus1}
\end{equation}
where
\begin{equation}
\langle A(t) \rangle_t =\frac{1}{T}\int_{t_0}^{t_0+T}A(t)
\label{sus2}
\end{equation}
is an average over an observation duration $T$. $t_0$ is an arbitrary time when the system reaches a steady state. If $J=K$, then the swarmalator model is reduced to two uncoupled Kuramoto models for $\zeta$ and $\eta$. In this case, using result of \cite{yoon2015critical}, we find explicit equations for the susceptibilities $\chi_{+}$ and $\chi_{-}$,
\begin{equation}
\chi_{\pm}= \begin{cases} \frac{1}{K_c-K}, &  K < K_c \\
\frac{K_c}{2K(K - K_c)}, & K > K_c
\end{cases}
\label{sus3}
\end{equation}
where $K_c=J_{+,c}=2(\Delta_v + \Delta_{\omega})$. At the critical point $K_c$, the susceptibility $\chi_{\pm}$ diverges. Note that $\chi_{\pm}$ demonstrates asymmetrical behavior below and above the critical point. \\

At $J\neq K$ we calculated the susceptibilities Eq. (\ref{sus1}) by use of numerical simulations of the microscopic Eqs. (1) and (2) of the main text. Figure \ref{fig:sus} shows the dependence of $\chi_{\pm}$ on the coupling $K$ at a constant $J=9$ (we used the same parameters as for Fig. 3(c) of the main text). The susceptibility $\chi_{\pm}$ first has a peak at $K=-1$ that corresponds to the critical point of the phase transition into the phase wave state, see Eq. (20) in the main text. This peak is an indicator of a phase transition from async state $(0,0)$ into the phase wave state $(S,0)$. With further increasing $K$, the susceptibility $\chi_{-}$ has one more peak at a critical value $K_{c}^{(m)} \approx 1.95$ that indicates a transition from the state $(S,0)$ into a state with non-zero $S_{-}~(=S_{2})$. It is the intermediate mixed state $(S_{1},S_{2})$ with $S_{1} > S_{2}$.
As we have shown in the Sec. IV, in this mixed state the corresponding functions, $\alpha$ and $\beta$, are time-dependent, see Eqs.~(\ref{eq: pws0}), (\ref{eq: pws16}) and (\ref{eq: pws17}). When increasing $K$ above $K_{c}^{(m)}$, the model transits into the sync state $(S,S)$ that is indicated by a cusp of $\chi_{+}$ at a critical value $K_c \approx 2.6$ that agrees with Eq. (23) in the main text. Equivalently, when increasing $K$ the swarmalator model can demonstrate a sequence of transitions: $(0,0) \mapsto (0,S)\mapsto (S_{1},S_{2}) \mapsto (S,S)$ where $S_{1} < S_{2}$.
\begin{figure}[htpb]
\centering
\includegraphics[width=8cm,angle=0.]{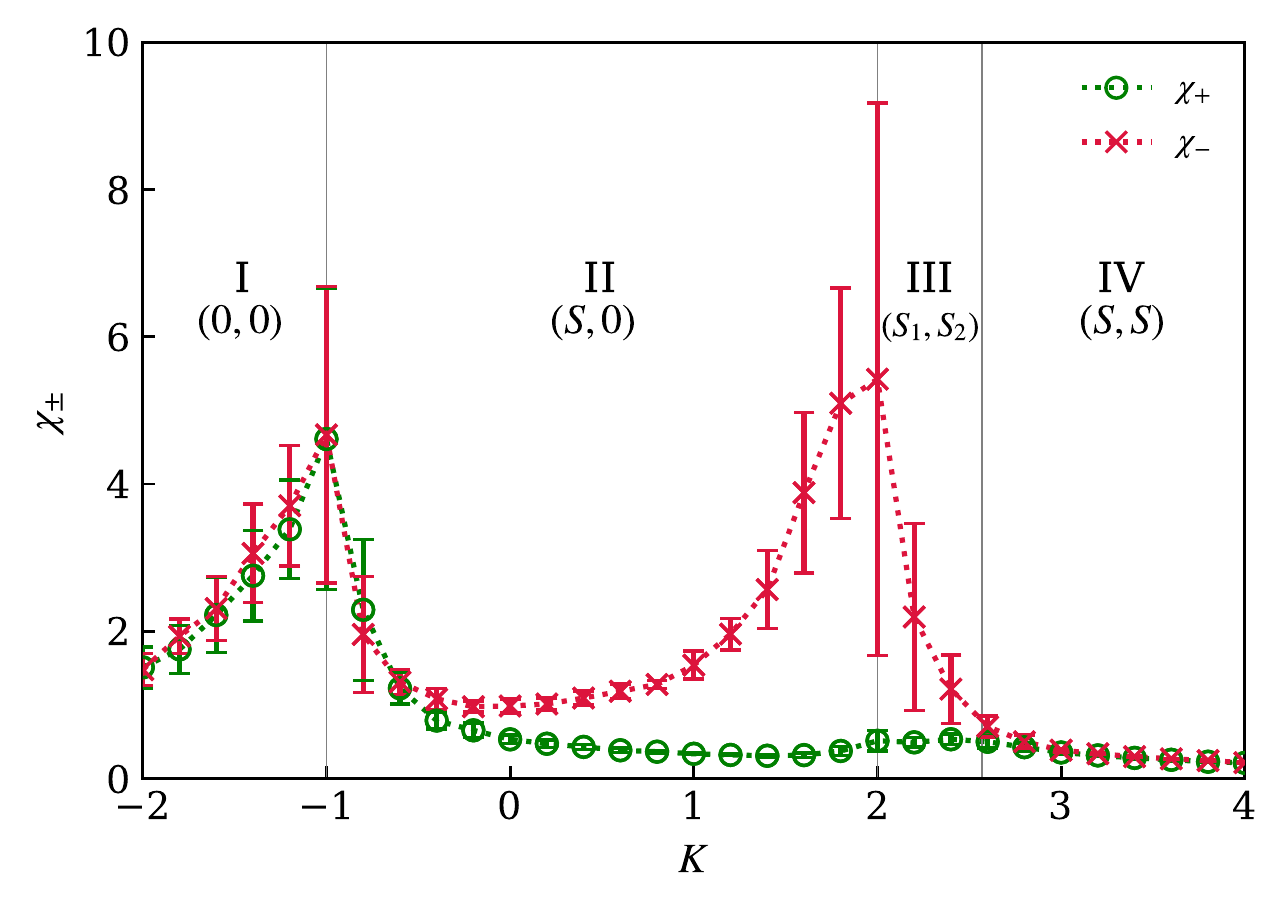}
\caption{ Susceptibility $\chi_{\pm}$ versus the coupling $K$ in the swarmalator model at the couplings $J=9$. Other parameters: the number of swarmalator $N=10^4$, the observation time $T=1000$, the spreads $\Delta_v=\Delta_{\omega}=1$, the initial time step $\delta t=0.01$ in the adaptive RK45.}
\label{fig:sus}
\end{figure}

\section{Sync state at $J=K$}
When $J = K \Rightarrow J_- = 0$, Eqs. (\ref{eq: 16_SM}) and (\ref{eq: 17_SM}) are uncoupled to 
\begin{eqnarray}
\frac{d \alpha}{d t}= - i(v+\omega) \alpha + \frac{1}{2}J (W^{\ast}_{+}- W_{+} \alpha ^2),
\label{eq: 30} \\
\frac{d \beta}{d t}=- i(v-\omega) \beta + \frac{1}{2}J (W^{\ast}_{-}- W_{-} \beta ^2),
\label{eq: 31}
\end{eqnarray}
and the phases $\zeta$ and $\eta$ evolve independently according to a regular Kuramoto model. Notice $\alpha(v,\omega,t)$ and $\beta(v,\omega,t)$ depend on $v+\omega$ and $v-\omega$, respectively, which are both distributed according to a Lorentzian with spread $\Delta_{v}+\Delta_{\omega}$. This allows the integrals for $W_{\pm}$, Eqs.~$(\ref{eq: 18_SM})$,$(\ref{eq: 19_SM})$, to be computed explicitly: $W_{+}(t)=\alpha^{\ast}(i\Delta_{v}+i\Delta_{\omega},t)$ and $W_{-}(t)=\beta^{\ast}(i\Delta_{v}+i\Delta_{\omega},t)$. Plugging these into Eqs. (\ref{eq: 30}) and (\ref{eq: 31}) yields
\begin{equation}
\frac{d W_{\pm}}{d t}= - (\Delta_{v}+\Delta_{\omega}) W_{\pm} + \frac{1}{2}J_+ (W_{\pm}- \| W_{\pm} \|^2 W_{\pm}).
\label{eq: 32}
\end{equation}
These equations are identical to those of the regular Kuramoto model \cite{ott2008low,ott2009long,childs2008stability}, as expected. The steady state order parameters are the same as that of the phase wave: $S_+ = S_- = S$ where $S$ is given by Eq. (19) with $J=K$ in the main text. The bifurcation line is $J=K$. As shown in Fig. 2(a) of the main text, this implies it is possible to pass from the async to the sync state without passing through the phase wave.

\section{Microscopic analysis of sync state}
Here, from Eqs. (\ref{xi_eom}) and (\ref{eta_eom}), the locked swarmalators obey $\dot{\zeta}_i = \dot{\eta}_i = 0$ which imply
\begin{eqnarray}
    \sin \zeta_i^* = \frac{K v + J \omega}{JK S_+}
    \label{eq: sync_micro1} \\
    \sin \eta_i^* = \frac{K v - J \omega}{JK S_-}
    \label{eq: sync_micro2}
\end{eqnarray}
where $\Phi_{\pm} = 0$ wlog. Self-consistency requires
\begin{equation}
    S_+ =  \int_{\Gamma} \cos(\zeta) g(\omega) h(v) d \omega dv
    \label{eq: sync_micro3}
\end{equation}
where $\Gamma$ denotes the locked swarmalator region in $(\omega, \nu)$ space, $-1 \leq \sin \zeta^* \leq 1 \cup -1 \leq \sin \eta^* \leq 1$. $S_-$ obeys a similar equation.  \\

Surprisingly, this seemingly simple and standard approach contradicts to our numerical simulations, the explicit equations Eq. (19) of the main text and Eq. (\ref{eq: 32}),  because it does not explain the critical boundary between the phase wave and the sync states. Instead of the observed continuous transition it predicts a discontinuous transition. There must be a nontrivial solution of microscopic equations (\ref{xi_eom}) and (\ref{eta_eom}) to explain the transition into the sync state. The explicit OA ansatz may help to solve this problem. In future work  we hope to better explore this issue.
\section{Verify generalized OA ansatz on double delta model}
Here we test the generalized OA ansatz against a different `double delta' frequency distribution: $g_{\omega}(\omega) = \frac{1}{2}\delta(\omega-\Delta) + \frac{1}{2} \delta(\omega + \Delta)$ and $g_{\nu}(\nu) = \frac{1}{2}\delta(\nu - \Delta) + \frac{1}{2} \delta(\nu + \Delta)$. Note, like the regular OA ansatz, the generalized OA ansatz \cite{ott2008low} does \textit{not} hold when the oscillator are precisely identical $\Delta = 0$, but does hold in the \textit{limit} $\Delta \rightarrow 0$.  \\

Simulations show this instance of the model has several collective states: (i) full sync; all swarmalators locked and $S_{\pm}$ fixed (ii) async (all swarmalator drifting) $S_{\pm} = 0$ (iii) non-stationary states.  Since our goal is simply to verify the ansatz, we analyze the sync state only, and leave the other collective states for future explorations.  \\

In the sync state, the order parameters $S_+$ is given by Eq.(18) in the main text, where the function $H^{\ast}\Bigl(\frac{v+\omega}{S_{+}J_{+}} \Bigr)$ is replaced to $H^{\ast}\Bigl(\frac{v}{JS_{+}}{+}\frac{\omega}{KS_{+}} \Bigr)$ according to Eq. (21) of the  main text, 
\begin{equation}
S_{+}= \int_{-\infty}^{\infty} dv \int_{-\infty}^{\infty} d\omega g_{v}(v) g_{\omega}(\omega) 
H^{\ast}\Bigl(\frac{v}{JS_{+}}{+}\frac{\omega}{KS_{+}} \Bigr),
\end{equation}
where,
\begin{equation}
    H(x) = -i x + \sqrt{1-x^2}
\end{equation}
(The same equation holds for $S_-$; that is, $S_{+} = S_{-} := S$). Subbing the double delta $g_{\omega}, g_{\nu}$ into this yields the following implicit equation for $S_{+}$
\begin{equation}
S_{+}= \frac{1}{4} \sum_{j=1}^4 \sqrt{ 1 - \Big( \frac{\Delta_j}{S_+} \Big)^2}
\end{equation}
where 
\begin{align}
\Delta_1 &= \Delta \Big( -\frac{1}{J} - \frac{1}{K} \Big) \\
\Delta_2 &= \Delta \Big( -\frac{1}{J} + \frac{1}{K} \Big) \\
\Delta_3 &= \Delta \Big( \frac{1}{J} - \frac{1}{K} \Big) \\
\Delta_4 &= \Delta \Big( \frac{1}{J} + \frac{1}{K} \Big) 
\end{align}

For $\Delta = 1$  (which can be achieved by rescaling time), Mathematica finds an exact (albeit complex!) solution to this equation:
\begin{align}
    S_+^2 &= -\frac{1}{2} \sqrt{\frac{\sqrt[3]{2} A}{3 J^2 K^2 \sqrt[3]{\sqrt{B^2-4 A^3}+B}}+\frac{\sqrt[3]{\sqrt{B^2-4 A^3}+B}}{3 \sqrt[3]{2} J^2 K^2}-\frac{2 \left(J^2+K^2\right)}{3 J^2 K^2}+\frac{1}{4}} \\
    & -\frac{1}{2}
   \sqrt{\frac{1}{4} -\frac{A}{3 J^2 K^2 \sqrt[3]{\sqrt{B^2-4 A^3}+B}}-\frac{1-\frac{4 \left(J^2+K^2\right)}{J^2 K^2}}{4 \sqrt{\frac{A}{3 J^2 K^2 \sqrt[3]{\sqrt{B^2-4 A^3}+B}}+\frac{\sqrt[3]{\sqrt{B^2-4 A^3}+B}}{3
   \sqrt[3]{2} J^2 K^2}-\frac{2 \left(J^2+K^2\right)}{3 J^2 K^2}+\frac{1}{4}}}} \\ 
   &-\frac{\sqrt[3]{\sqrt{B^2-4 A^3}+B}}{3 \sqrt[3]{2} J^2 K^2}-\frac{4 \left(J^2+K^2\right)}{3 J^2 K^2}+\frac{1}{2} \label{S_th}
\end{align}
where
\begin{align}
    A &= 14 J^2 K^2+J^4+K^4 \\
    B &= 27 J^4 K^4-72 J^2 K^2 \left(J^2+K^2\right)+2 \left(J^2+K^2\right)^3
\end{align}
In order for $S$ to be real, we require $B^2 - 4A^3 > 0$ which means the sync state is born at
\begin{align}
\left(27 J^4 K^4-72 J^2 K^2 \left(J^2+K^2\right)+2 \left(J^2+K^2\right)^3\right)^2-4 \left(14 J^2 K^2+J^4+K^4\right)^3 = 0 \label{Kc}
\end{align}
Figure~\ref{fig:S_th_test}(a) below shows this theoretical expressions for $S_+$ and $K_c(J,\Delta)$ agrees with simulations perfectly, proving the generalized OA ansatz is correct. Note, the erratic red dots correspond to an unsteady state in which $W_{\pm}$ oscillate. We intend to explore this state and its bifurcations (we suspect it is a SNIC) as well as the other states of the model in future work. Figure~\ref{fig:S_th_test}(b) shows the relaxation of $S_+$  which also agrees well with simulation.The Mathematica notebook containing this analysis is available at \footnote{https://github.com/Khev/swarmalators/blob/master/1D/on-ring/non-identical/big-N-sim.nb}.

\begin{figure}
    \centering
    {{\includegraphics[width=7.2cm]{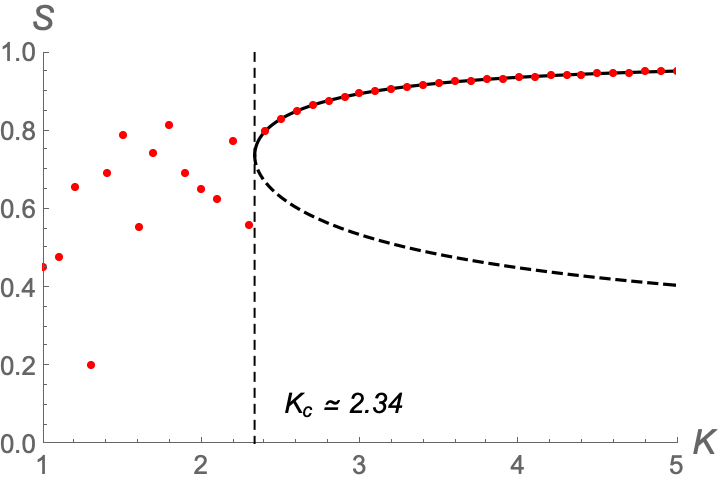} }}%
    \qquad
    {{\includegraphics[width=8cm, height=4.8cm]{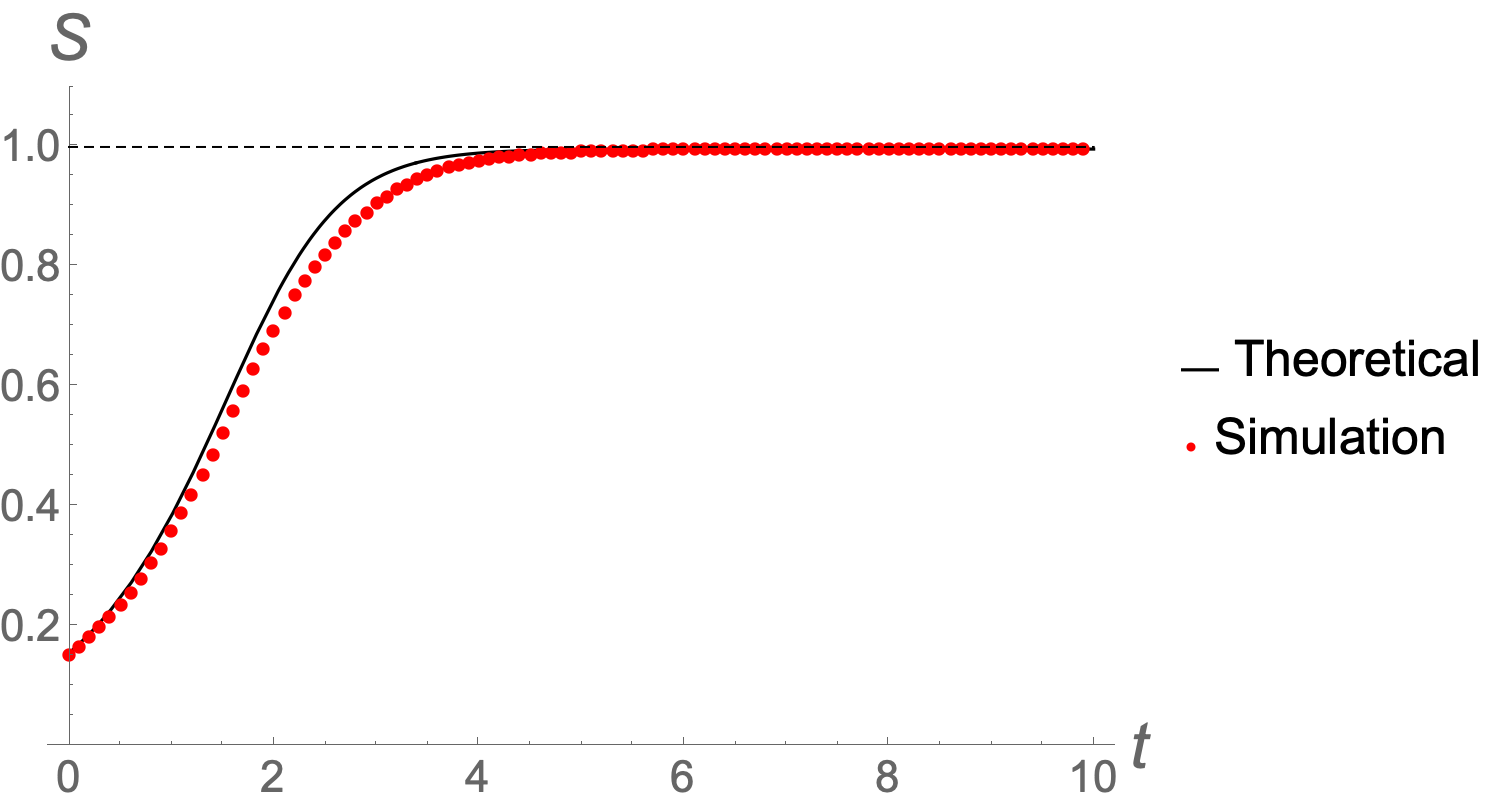} }}%
    \caption{(a) Theoretical prediction of $S_+ := S$ Eq.~\eqref{S_th} (solid black curve) and $K_c \approx 2.34$ (derived from solving Eq.~\eqref{Kc})  agree perfectly with simulation (red dots; averaging over final $10$ to avoid transients). The unstable branch of $S_{+}$ is plotted as a dashed curve and shown to illustrate the saddle node bifurcation. Notice for $K < K_c$, the system is in a non-stationary state as indicated by the erratic red dots (this state is not analyzed here; see text above). Simulation details: $N = 100$ swarmalators integrated with an RK4 solver with $(dt, T) = (0.1, 2000)$. Parameters: $(J, \Delta) = (5,1)$ (b) Time evolution of $S$. Simulation details: $N = 200$ swarmalators integrated with an RK4 solver with $(dt, T) = (0.1, 2000)$. Parameters: $(J, \Delta) = (1, 0.25)$ }%
    \label{fig:example}%
\end{figure}

Note that the double delta model demonstrates a phase transition from a phase with oscillations of the order parameters $W_{\pm }$ into the sync state.
\section{Connection of ring model to a 2D swarmalator model}
Here we show how the ring model is contained within the 2D swarmalator model which is given by
\begin{align}
&\dot{\mathbf{x}}_i = \mathbf{v}_i + \frac{1}{N} \sum_{j=1}^N \Big[ \mathbf{I}_{\mathrm{att}}(\mathbf{x}_j - \mathbf{x}_i)F(\theta_j - \theta_i)  - \mathbf{I}_{\mathrm{rep}} (\mathbf{x}_j - \mathbf{x}_i) \Big],   \\
& \dot{\theta_i} = \omega_i +  \frac{K}{N} \sum_{j=1}^N H_{\mathrm{\mathrm{att}}}(\theta_j - \theta_i) G_{\sigma}(\mathbf{x}_j - \mathbf{x}_i)
\end{align}
In \cite{o2017oscillators}, the choices $I_{att} = x / |x|$, $I_{rep} = x / |x|^2$, $F(\theta) = 1 + J \cos(\theta)$, $G(x) = 1 / |x|$, $H_{att}(\theta) = \sin(\theta)$ were made. However, choosing linear spatial attraction $I_{att}(x) = x$, inverse square spatial repulsion $I_{rep}(x) = x / |x|^2$ and truncated parabolic space-phase coupling $G(x) = (1 - |x|^2/ \sigma^2) H_{heaviside}({\sigma - |x|})$
\begin{align}
&\dot{\mathbf{x}}_i = \frac{1}{N} \sum_{ j \neq i}^N \Bigg[ \mathbf{x}_j - \mathbf{x}_i \Big( 1 + J \cos(\theta_j - \theta_i)  \Big) -   \frac{\mathbf{x}_j - \mathbf{x}_i}{ | \mathbf{x}_j - \mathbf{x}_i|^2}\Bigg] \label{linear_parabolic1} \\
& \dot{\theta_i} = \frac{K}{N} \sum_{j \neq i}^N \sin(\theta_j - \theta_i ) \Big(1 - \frac{| \mathbf{x}_j - \mathbf{x}_i|^2}{\sigma^2} \Big) H_{heaviside}(\sigma - |\mathbf{x}_j - \mathbf{x}_i|) \label{linear_parabolic2}
\end{align}
\noindent
gives the same qualitative behavior but is nicer to work with analytically. We call this the `linear parabolic` model because $I_{att} = x$ and $G(x)$ is a parabolic, In polar coordinates it takes form
\begin{align*}
\dot{r_i}&=  H_r(r_i, \phi_i) - J r_i R_0 \cos \Big( \Psi_0 - \theta_i \Big) + \frac{J}{2} \Bigg[ \tilde{S}_+ \cos \Big( \Phi_+- (\phi_i+\theta_i) \Big)
 +  \tilde{S}_{-}\cos \Big( \Phi_{-} - (\phi_i-\theta_i) \Big) \Bigg] \\
\dot{\phi_i}&=  H_{\phi}(r_i, \phi_i)  + \frac{J}{2 r_i} \Bigg[ \tilde{S}_+ \sin \Big( \Psi_+ - (\phi_i+\theta_i) \Big)  + \tilde{S}_- \sin \Big( \Psi_{-} -  (\phi_i- \theta_i) \Big) \Bigg] \\
\dot{\theta_i} &=  K \Big(1- \frac{r_i^2}{\sigma^2} \Big) R_0 \sin (\Phi_0 - \theta_i) - \frac{K}{\sigma^2} R_1 \sin(\Phi_1 - \theta_i) + \frac{K r_i}{\sigma^2} \Bigg[ \tilde{S}_+ \sin \Big( \Psi_+ - (\phi_i+\theta_i) \Big) - \tilde{S}_- \sin \Big( \Psi_- - (\phi_i- \theta_i) \Big) \Bigg]
\end{align*}
\noindent
where
\begin{align}
H_r(r_i, \phi_i) &=  \frac{1}{N} \sum_{j} \Big( r_j \cos(\phi_j - \phi_i) - r_i \Big) ( 1 - d_{ij}^{-2} )  \label{Hr} \\
H_{\phi}(r_i, \phi_i) &=  \frac{1}{N} \sum_{j} \frac{r_j}{r_i} \sin(\phi_j - \phi_i) ( 1 - d_{ij}^{-2} ), \label{Hphi}  \\
Z_0 = R_0 e^{i \Psi_0}&=  \frac{1}{N} \sum_{j}  e^{i \theta_j},
\\
\hat{Z}_0 = \hat{R}_0 e^{i \hat{\Psi}_0}&=  \frac{1}{N} \sum_{j \in N_i}  e^{i \theta_j}, \label{Qzero} \\
Z_2 = R_2 e^{i \Psi_2}&=  \frac{1}{N} \sum_{j} r_j^2 e^{i \theta_j}, \label{Q1} \\
\hat{Z}_2 = \hat{R}_2 e^{i \hat{\Psi}_2}&=  \frac{1}{N} \sum_{j \in N_i} r_j^2 e^{i \theta_j}, \label{Q2} \\
\tilde{W}_{\pm} = \tilde{S}_{\pm} e^{i \Psi_{\pm}}&=  \frac{1}{N} \sum_{j} r_j e^{i (\phi_j \pm \theta_j)}
\label{Qplusminus} \\
\hat{W}_{\pm} = \hat{S}_{\pm} e^{i \hat{\Psi}_{\pm}}&=  \frac{1}{N} \sum_{j \in N_i} r_j e^{i (\phi_j \pm \theta_j)}
\end{align}
\noindent
where the $\hat{Z_0}, \dots$ order parameters are summed over all the neighbours $N_i$ of the $i$-th swarmalator: those within a distance $\sigma$. Notice that rainbow order parameters $\tilde{W}$ here are weighted by the radial distance $r_j$, which is not the case for the ring model presented in the text( that's why we put a tilde over the W).  Assuming $\sigma > max(d_{ij})$, we can set $\hat{Z_0} = Z_0, \hat{Z}_1 = Z_1, \hat{W_\pm} = W_{\pm}$. If we assume there is no global synchrony $Z_0 = Z_2 = 0$, which happens generically in the frustrated parameter regime $K < 0, J > 0$, and transform to $\zeta_i = \phi_i + \theta_i$ and $\eta_i = \phi_i - \theta_i$ coordinates the ring model is revealed (the terms in the square parentheses in the latter two equations.)
\begin{align}
\dot{r_i}&= \tilde{\nu}(r_i) +  \frac{J}{2} \Bigg[ \tilde{S}_+ \cos \Big( \Phi_+- \zeta_i \Big) + \tilde{S}_{-}\cos \Big( \Phi_{-} - \eta_i \Big) \Bigg] \\
\dot{\zeta_i}&= \tilde{\omega}(r_i, \phi_i) + \Bigg[ J_+(r_i) \tilde{S}_+ \sin \Big( \Psi_+ - \zeta_i \Big)  + J_- (r_i) \tilde{S}_- \sin \Big( \Psi_{-} -  \eta_i \Big) \Bigg]   \\
\dot{\eta_i} &= \tilde{\omega}(r_i, \phi_i) +  \Bigg[ J_-(r_i) \tilde{S}_+ \sin \Big( \Psi_+ - \zeta_i \Big) - J_+(r_i) \tilde{S}_- \sin \Big( \Psi_- - \eta_i \Big) \Bigg]
\end{align}
where
\begin{align}
\tilde{\nu}(r_i, \phi_i) &= H_r(r_i, \phi_i) \\
\tilde{\omega}(r_i, \phi_i) &= H_{\phi}(r_i, \phi_i) \\
J_{\pm}(r_i) &= \frac{J}{2 r_i} \pm \frac{K r_i}{\sigma^2}
\end{align}

which has the same form as the ring model presented in the paper; in that sense the ring model captures an aspect of the 2D swarmalator model's rotational motion.

\section{Stability and sizes of clusters of locked swarmalators}

As we showed in the main text, the swarmalator model, which is determined by equations (1) and (2), obeys an internal symmetry with respect to the $\pi$-transformation. Namely, a replacement of the phases $x_i$ and $\theta_i$ of an arbitrary $i$-th swarmalator to $\tilde{x}_i = x_i +\pi$ and $\tilde{\theta}_i= \theta_i +\pi$ does not change these equations. It means that a transition of the $i$-th swarmalator $i$ from a state $(\theta_i,x_i)$ into a `mirrored' state $(\theta_i+\pi, x_i +\pi)$ does not influence on dynamics of other swarmalators. This symmetry allows the formation of stable mirrored clusters of synchronized swarmalators.

\begin{figure}[htpb!]
\centering
\includegraphics[width=12cm,angle=0.]{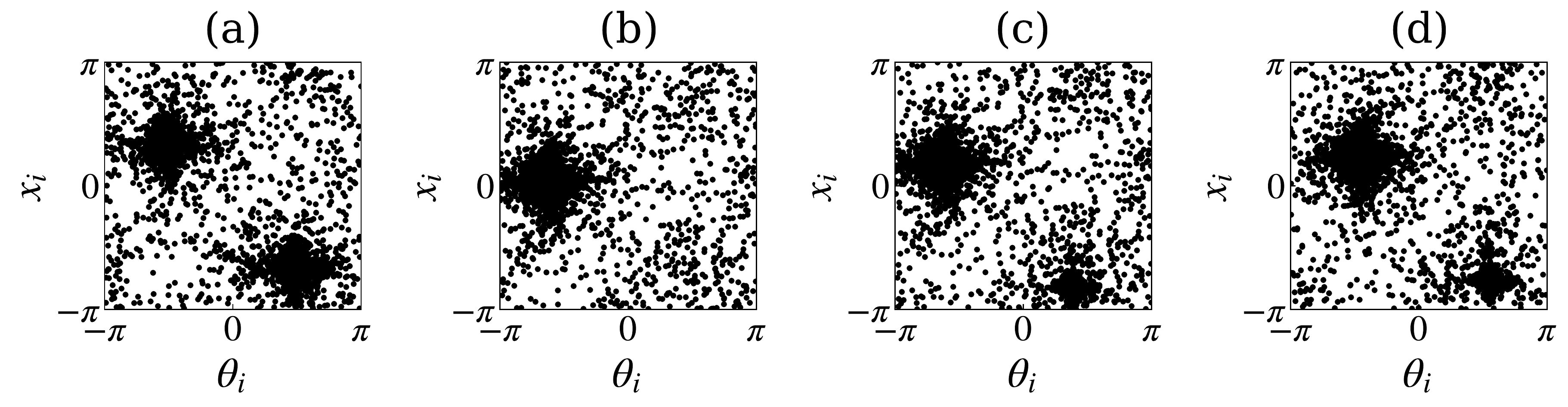}
\caption{Snapshots of the sync state $(S,S)$ in the $(\theta,x)$ plane at different initial conditions. (a) Initial phases of the swarmalators are chosen at random in the interval $[0,2\pi)$. We observed two mirrored clusters of swarmalators with locked phases $\theta_i$ and $x_i$. (b) Uniform initial conditions, $\theta_i=x_i=\pi$, for $i=1,2, \dots , N$. We found one cluster of locked swarmalators. (c) The sync state in the panel (b) after application of the $\pi$-transformation with the probability $p=1/10$ to the locked swarmalators, for details, see the text.
(d) The state in the panel (c) after the observation time $T=2000$. Parameters: the number of swarmalator $N=5000$, the couplings $K=J=10$, the observation time $T=2000$, the spreads $\Delta_v=\Delta_{\omega}=1$, the initial time step $\delta t=0.01$ in the adaptive RK45 algorithm.}
\label{fig:mc}
\end{figure}

In order to demonstrate these properties, we performed a numerical solution of the Eqs. (1) and (2) at parameters corresponding to the steady sync state $(S,S)$. Snapshots of swarmalators' phases in the $(\theta,x)$ plane at different initial conditions are represented in Figs. \ref{fig:mc}(a) and \ref{fig:mc}(b). In the case when the swarmalators´ phases were chosen at random in the interval $(0,2\pi]$, in the steady state we observed two mirrored clusters of swarmalators with locked phases  $\theta_i$ and $x_i$, see the Fig. \ref{fig:mc}(a). They were of approximately the same size. In the case of the uniform initial conditions, $\theta_i=x_i=\pi$ for all $i=1, 2, \dots , N$, we found only one cluster of swarmalators with locked phases in the steady state, see the Fig. \ref{fig:mc}(b). Then we applied the $\pi$-transformation to swarmalators in this cluster. Namely, these swarmalators are moved by the $\pi$-transformation with the probability $p$ to the mirrored cluster. Thus, we formed a mirrored cluster that had got approximately the size $pN_{ls}$, where $N_{ls}$ is the  size of the initial cluster on the panel Fig. \ref{fig:mc}(b). A snapshot of the obtained state is shown on Fig. \ref{fig:mc} (c). Using this state as an initial condition in the Eqs. (1) and (2), we found that this state is stable and does not change in time, see a snapshot on the Fig. \ref{fig:mc}(d).

\begin{figure}[htpb!]
\centering
\includegraphics[width=6cm,angle=0.]{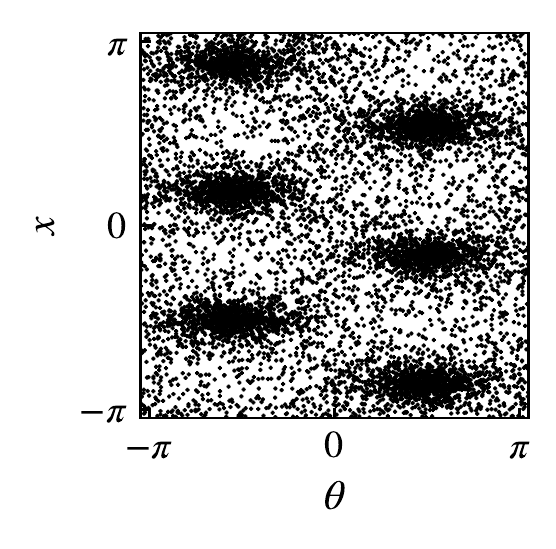}
\caption{Snapshots of the sync state $(S,S)$ in the $(\theta,x)$ plane in the case of the length scale $L_x =3$ in the coupling term. Initial phases of the swarmalators are chosen at random in the interval $[0,2\pi)$. There are 6 clusters of swarmalators with locked phases $\theta_i$ and $x_i$. Parameters: the number of swarmalator $N=10^4$, the couplings $K=J=5$, the observation time $T=200$, the spreads $\Delta_v=\Delta_{\omega}=1$, the initial time step $\delta t=0.01$ in the adaptive RK45 algorithm.}
\label{fig:mc2}
\end{figure}

There are multiple clusters of locked swarmalators in the swarmalator model with a length scale $L_x$ in the coupling term, namely, $L_x (x_j -x_i)$. In this case, the microscopic Eqs. (1) and (2) have a form
\begin{eqnarray}
\dot{x}_i=v_i+\frac{J}{N} \sum_{j=1}^N \sin[L_x (x_j - x_i)] \cos(\theta_j-\theta_i),
\label{eq: 1} \\
\dot{\theta}_i =  \omega_i+\frac{K}{N} \sum_{j=1}^N \sin(\theta_j-\theta_i) \cos[L_x (x_j - x_i)],
\label{eq: 2}
\end{eqnarray}
For simplicity we consider the case when $L_x$ is an integer number. A replacement of the phases $x_i$ and $\theta_i$ of an arbitrary $i$-th swarmalator to $\tilde{x}_i = x_i +l\pi /L_x$ and $\tilde{\theta}_i= \theta_i +\pi$, where $l$ is a positive or negative odd number, $|l| \leq L_x$, does not change these equations. The total number of equivalent clusters is $2L_x$. Note that the number of swarnalators in each cluster is a random number but the total number of locked swarmalators in these clusters, is fixed and determined by the couplings and related with the order parameters $S_{\pm}$. Distribution of the locked swarmalators over clusters depends on initial conditions as we have discussed above for the case $L_x =1$. Figure \ref{fig:mc2} shows that in the case of $L_x =3$, in the sync state the locked swarmalators forms 6 clusters with almost equal density of swarmalators because initial phases $x_i$ and $\theta_i$ were chosen at random. One also can add a scale length $L_{\theta}$ in the coupling $\theta_j -\theta_i$ that  increases the total number of clusters.

\section{Local coupling}
\begin{figure}[htpb!]
\centering
\includegraphics[width=13 cm,angle=0.]{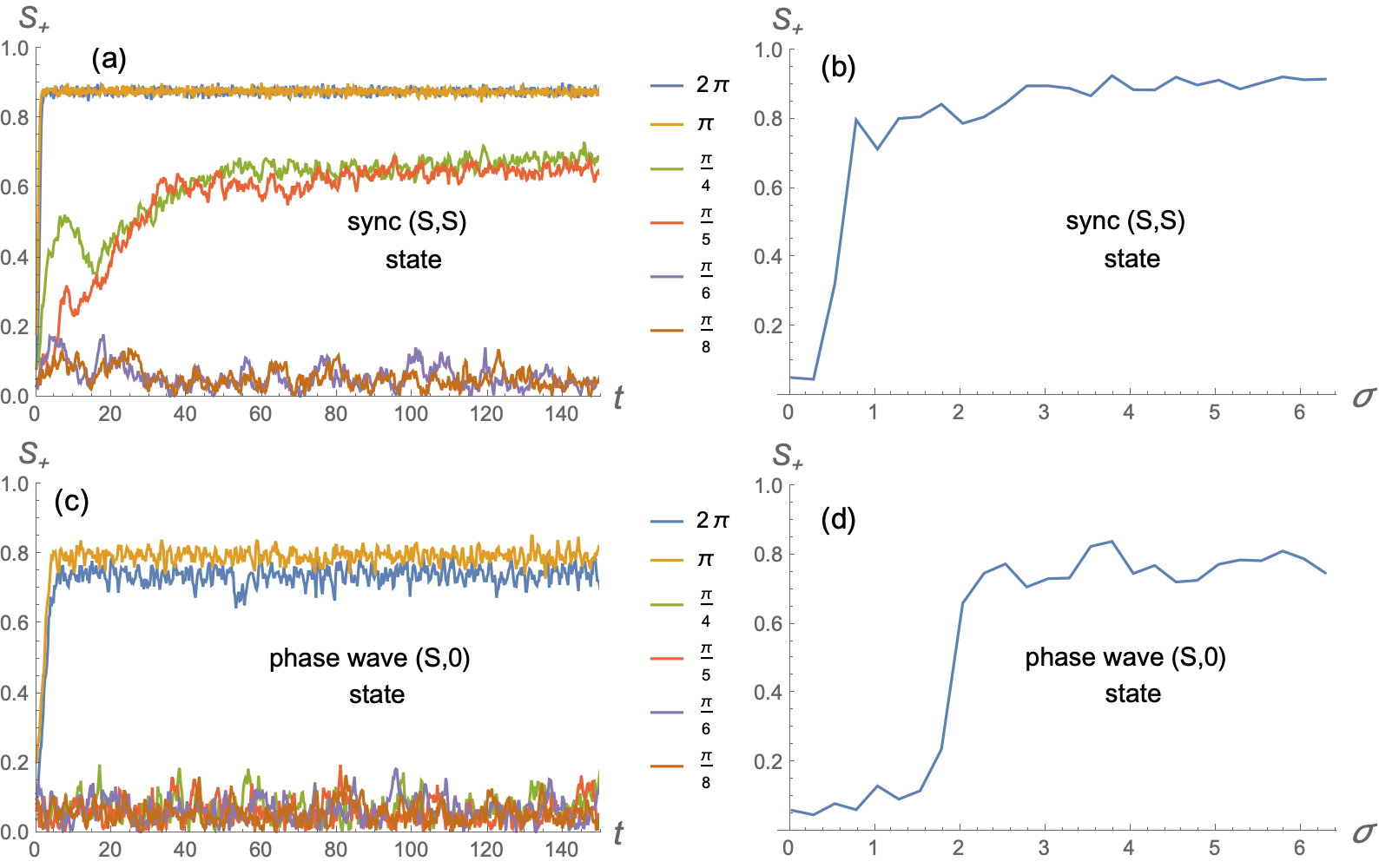}
\caption{Order parameters for local coupling. First column shows time series of $S_+$ (where we assume wlog that $S_+ > S_-$). Second column shows the steady state value of $S_+$ versus $\sigma$. Simulation details: RK4 method with $(dt, T, N) = (0.25, 200, 500)$. Top row: $(S,S)$ sync state, 
$(J,K,\Delta) = (5,7,0.25)$. Bottom row, $(S,0)$ phase wave $(J,K,\Delta) = (5,-0.1,0.25)$. }
\label{fig:local_coupling}
\end{figure}

Our model has mean-field or global coupling, where each element interacts with every other element of the population. Most active matter systems, however, have local coupling, where each element can only sense its local neighbours and thus interacts with a subset of the population. We were curious how well our mean field model approximated a model with local modeling limits so we added a finite cutoff with range $\sigma \in [0, 2 \pi]$ to our model,
\begin{eqnarray}
\dot{x}_i=v_i+\frac{J}{N} \sum_{j=1}^N \sin(x_j - x_i) \cos(\theta_j-\theta_i) H(\sigma - |x_j-x_i|_{geo}),
\label{eq: local_coupling_1} \\
\dot{\theta}_i =  \omega_i+\frac{K}{N} \sum_{j=1}^N \sin(\theta_j-\theta_i) \cos(x_j - x_i) H(\sigma - |x_j-x_i|_{geo}),,
\end{eqnarray}
where $H(x)$ is heaviside's step function and the distance $|x_j - x_i|_{geo} := \min( abs(x_j -x_i), abs(x_j - x_i + 2\pi), abs(x_j - x_i - 2 \pi))$ is the geodesic distance between $x_j, x_i$. We consider the geodesic distance, because, recall, $x_1, x_2$ are angles on the unit circle.  \\

A full study of the model above for all $(J,K,\Delta, \sigma)$ is out of scope. We concern ourselves with the validity of the mean field approximation, namely, the robustness of the collective states (sync, phase wave) to increasing locality (decreasing $\sigma$). Figure~\ref{fig:local_coupling} shows the results. The sync and phase wave persist for $\sigma \gtrsim \pi$, then gradually blur (the blur corresponds to smaller $S_+$) until finally disappearing. The sync state becomes the async state (although transient states with multiple clusters were observed), but the phase wave morphs into a series of phases waves with winding number $k > 1$ (Figure~\ref{fig:local_coupling_phase_wave}) until finally becoming the async state.\\

These results of our numerical simulations presented on Figs. S5 and S6 demonstrate that the cutoff for $\sigma \gtrsim \pi$ in the spatial interaction kernel does not qualitatively change the dynamics of the swarmalator model.

\begin{figure}[htpb!]
\centering
\includegraphics[width=12 cm,angle=0.]{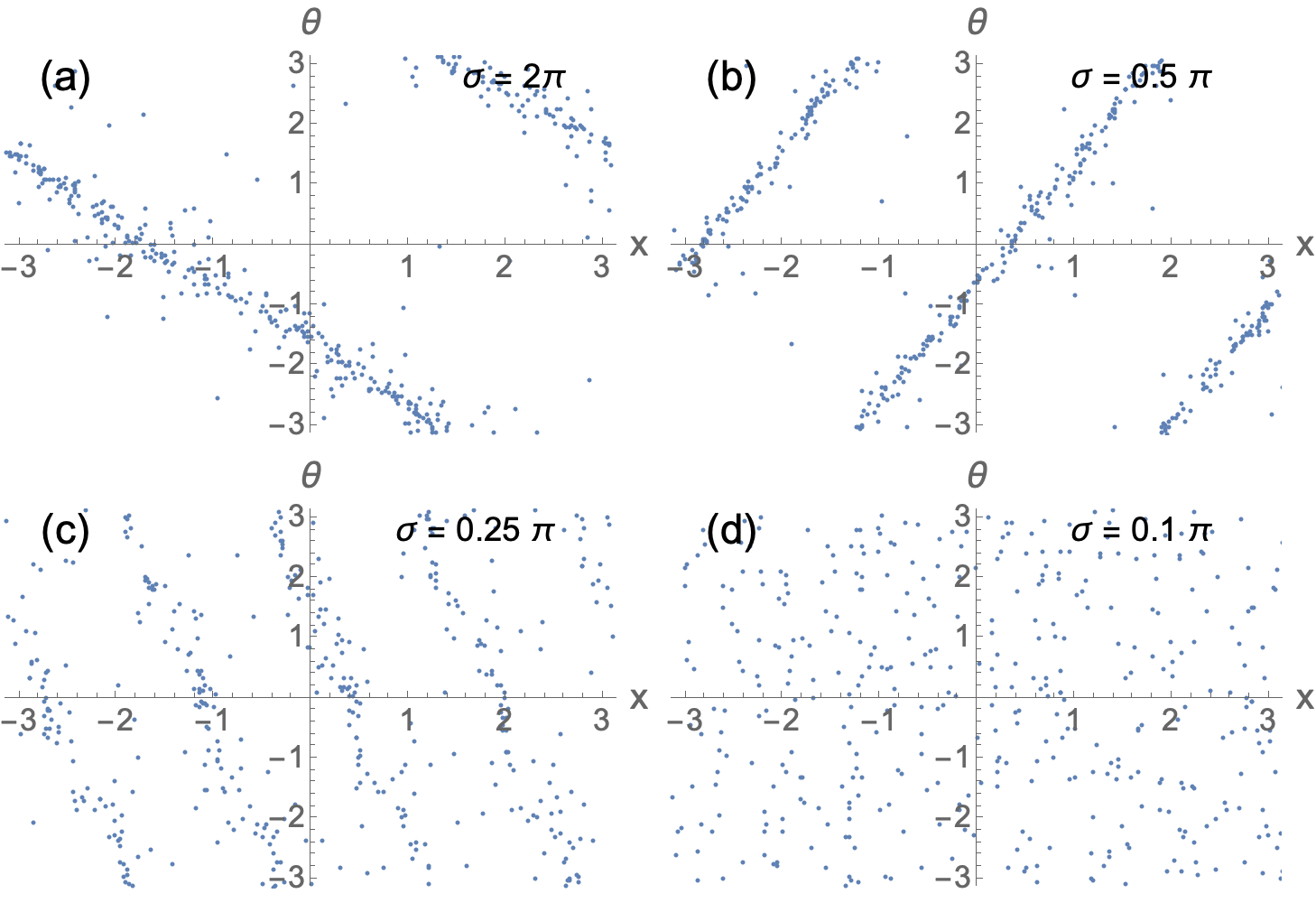}
\caption{Phase waves with different winding numbers $k$. Simulation details: RK4 method with $(dt, T, N) = (0.25, 200, 500)$ with $(J,K,\Delta) = (5,-0.1,0.25)$. (a) $\sigma = 2\pi, k=1$. (b) $\sigma = 0.5 \pi,  k=2$. (c) $\sigma = 0.25 \pi, k=4$. (d)  $\sigma = 0.25 \pi$, async state.}
\label{fig:local_coupling_phase_wave}
\end{figure}

\section{Movies}
We made movies of the four collective states by integreating the governing equations using an RK4 solver with $(dt, T) = (0.05, 25)$ for $N = 900$ swarmalators. The parameters for each state were the same as those in Figure 1 and Figure 2 in the main text (also quoted below). The initial positions and phases were drawn uniform from $[0,2\pi]$ in all cases except for the single cluster sync state, which were drawn from $[0, 0.5 \pi]$. The Lorentzian $g(\omega)$ were drawn  `deterministically': we prepared a set of variables $u_i$ linearly spaced on $[0,1]$ and then projected these using the inverse CDF of the Lorentzian $\omega_i = CDF^{-1}(u_i)$. The purpose was to enforce $\bar{\omega} = \bar{\nu} = 0$; if we drew them randomly, then $\bar{\omega}, \bar{\nu}$ were never quite zero, due to finite effects, which led to an unphysical (in the sense of a finite effect) drift in the phases of the order parameters $\Phi_{\pm}$.

\begin{itemize}
    \item Sync (two cluster): $(J,K,\Delta) = (8,9,1)$
    \item Sync (one cluster): $(J,K,\Delta) = (8,9,1)$
    \item Phase wave: $(J,K,\Delta) = (1,40,1)$
    \item Async: $(J,K,\Delta) = (1,1,1)$
    \item Mixed: $(J,K, \Delta) = (2,30,1)$
\end{itemize}

\bibliography{ref.bib}

\end{document}